\documentclass[fleqn,usenatbib]{mnras}
\usepackage{placeins}
\usepackage{newtxtext,newtxmath}

\usepackage[T1]{fontenc}

\DeclareRobustCommand{\VAN}[3]{#2}
\let\VANthebibliography\thebibliography
\def\thebibliography{\DeclareRobustCommand{\VAN}[3]{##3}\VANthebibliography}

\usepackage{tikz,xcolor,hyperref}

\definecolor{lime}{HTML}{A6CE39}
\DeclareRobustCommand{\orcidicon}{%
	\begin{tikzpicture}
	\draw[lime, fill=lime] (0,0) 
	circle [radius=0.16] 
	node[white] {{\fontfamily{qag}\selectfont \tiny ID}};
	\draw[white, fill=white] (-0.0625,0.095) 
	circle [radius=0.007];
	\end{tikzpicture}
	\hspace{-2mm}
}

\foreach \x in {A, ..., Z}{\expandafter\xdef\csname orcid\x\endcsname{\noexpand\href{https://orcid.org/\csname orcidauthor\x\endcsname}
			{\noexpand\orcidicon}}
}

\usepackage{graphicx}

\usepackage{ulem}

\newcommand{\ha}{H\,$\alpha$} 
\newcommand{\hb}{H\,$\beta$}

\usepackage{graphicx}	
\usepackage{amsmath}	

\title[Analysis of IFS observations of the PN Hen~2-108]{Analysis of Integral Field Spectroscopy observations of the planetary nebula Hen 2-108 and its central star}

\author[B. L. Miranda Marques et al.]{Bárbara L. Miranda Marques$^{1,2\orcidA{}}$ \thanks{E-mail: barbaralmmarques@gmail.com (BLMM)},
Hektor Monteiro$^{1\orcidB{}}$,
Isabel Aleman$^{1\orcidC{}}$, 
Stavros Akras$^{3\orcidD{}}$, 
Helge Todt$^{4}$,
\newauthor{and Romano L. M. Corradi$^{5,6}$}
\\
$^{1}$Instituto de Física e Química, Universidade Federal de Itajubá, Av. BPS 1303, Pinheirinho, 37500-903, Itajubá, MG, Brazil\\
$^{2}$Divisão de Astrofísica, Instituto Nacional de Pesquisas Espaciais, Av. dos Astronautas 1758, São José dos Campos, 12227-010, SP, Brazil\\
$^{3}$Institute for Astronomy, Astrophysics, Space Applications and Remote Sensing, National Observatory of Athens, GR 15236 Penteli, Greece\\
$^{4}$ Institut of Physics and Astronomy, Universität Potsdam, Karl-Liebknecht-Str. 24/25, 14476, Potsdam, Germany \\
$^{5}$Instituto de Astrof\'{i}sica de Canarias, E-38205 La Laguna, Tenerife, Spain\\
$^{6}$GRANTECAN, Cuesta de San Jos\'{e} s/n, E-38712, Bre\~{n}na Baja, La Palma, Spain\\
}

\date{Accepted XXX. Received YYY; in original form ZZZ}

\pubyear{2015}

\begin{document}
\label{firstpage}
\pagerange{\pageref{firstpage}--\pageref{lastpage}}
\maketitle

\begin{abstract}
The study of planetary nebulae provides important constraints for many aspects of stellar and Galactic evolution. Hen~2-108 is a poorly known planetary nebula with a slight elliptical morphology and a peculiar central star (CS), which has defied classification. In this work, we present the first detailed integral field spectroscopic study of the planetary nebula Hen~2-108 and its CS. We provide spatially resolved flux maps for important emission lines, as well as diagnostic maps of extinction and electronic density and temperature. Physical conditions and chemical abundances were also calculated from the integrated spectrum. The analysis was also performed with the code {\sc satellite} which uses a distinct strategy to evaluate physical and chemical properties. Both {\sc satellite} and traditional procedure give consistent results, showing some variation in physical and chemical properties. We detect and measure a number of faint heavy element recombination lines from which we find a significant abundance discrepancy factor for O/H, and possibly for N/H. Pseudo 3D photoionization models were used to assist in the interpretation with results supporting the low-ionisation nature of this nebula, indicating a CS with $T_\textrm{eff}=$~40~kK and a shell structure. The spectrum of the CS has been analysed with a detailed model for expanding atmospheres to infer stellar parameters, finding that it is a [Of/WN8] type with $T_*=$~41.5~kK, making it a new addition to a small set ($\sim$20) of rare objects.

\end{abstract}

\begin{keywords}
planetary nebulae: general -- planetary nebulae: individual: Hen~2-108 -- ISM: abundances
\end{keywords}


\section{Introduction}

Planetary nebulae (PNe) are the result of the evolution of stars with masses between approximately 1 and 8 M$_{\odot}$. At the end of their evolution, these stars eject their outermost layers, forming the PN. The remnant stellar hot core ionizes and heats the surrounding matter. As the stars undergo nucleosynthesis processes that change their composition during their lives, they result in a significant chemical enrichment of the interstellar medium. \citep{Pottasch84a, Kwok00}.

Despite having a good grasp of the general processes that lead to the observed morphologies, the details of the physical processes behind the production of the different shapes and substructures of these objects are not yet fully understood. Differences in morphologies of PNe could be related to to the mass loss episodes from the progenitors, the interactions between the companions in binary systems or magnetic fields, and/or a combination of more than one mechanism in the same object. \citep[and references therein]{PNeReview2002,Osterbrock06,Maciel11}.

In the last decades, integral field spectroscopy (IFS) has become an important observational technique \citep{Mediavilla11}, especially in the study of extended objects such as ionized nebulae. The added spatial coverage is an advantage over traditional long slit strategies. IFS allows us to obtain spatially resolved physical and chemical information, such as kinematics and chemical abundances among other interesting diagnostics, on the extent of the object.

Planetary nebulae, being often resolved extended objects in the sky and with a diversity of morphologies, are obvious targets of studies with IFS. The first work using IFS of galactic PNe was published just over a decade ago by \citet{Tsamis08}. There is an increasing number of published works using this technique to address a range of topics in the last years. In addition to the PN gas, its central star (CS) can be studied with 3D spectroscopy data, as seen in the works of \citet{Ali16}, \citet{Basurah16} and \citet{Danehkar14}. \citet{Garcia-Rojas16} also used IFS data from VIMOS to study the PN abundance discrepancy problem. Works such as \citet{Ali19}, \citet{Akras20}, \citet{2016A&A...588A.106W,Walsh18}, and \citet{Walsh20} are examples of recent studies of PNe using IFS. 

The PN Hen~2-108 is an object with an apparently simple, slightly elliptical morphology, with few detailed studies about it \citep{Schwarz92, Stanghellini08,Pottasch11,Gorny14}. The first high quality images of Hen~2-108 were obtained using narrowband imaging by \citet{Schwarz92}. They identified a projected elliptical morphology in H$\alpha$ and [\ion{O}{iii}]~$\lambda$5007 narrowband filter images. Hen~2-108 is also classified as approximately circular or elliptical by \citet{Acker92} and \citet{Tylenda03}, who determined the dimensions 13.6$\times$12.3~arcsec for this nebula. In the literature, the distance to Hen~2-108 is found in the range of 1.7 to 4.6~kpc \citep{1982ApJ...260..612D,Cahn_etal_1992,Mendez92,vandeSteene_etal_1995,1995ApJS...98..659Z,2002ApJS..139..199P,Phillips_2005,2004MNRAS.353..589P,Stanghellini08,GonzalesSantamaria_etal_2019}. 

Hen~2-108 is believed to be a young, low-ionization nebula, with estimates of its central star temperature  ranging from 26~kK to 50~kK \citep{Kudritzki_etal_1997,Phillips_2003,Pauldrach04,Hultzsch07,Pottasch11}.
Infrared, optical and ultraviolet spectra were used by \citet{Pottasch11} to determine physical conditions, abundances and stellar parameters. 
\citet{Gorny14} classified Hen~2-108 CS as a very late (VL) type, due to the presence of the \ion{C}{iv} $\lambda5801$ and \ion{C}{iii} $\lambda5695$ lines in its spectrum. The same authors also identified \ion{He}{ii} $\lambda4686$ and \ion{C}{ii} $\lambda7235$ lines as stellar, classifying it as a weak emission-line CS (WELS). However, as mentioned in \cite{Basurah16} while studying four PNe with supposed WELS CSs, the spatially resolved spectroscopy shows that lines such as the \ion{C}{iv}  and \ion{N}{iii} in recombination are either distributed in the nebula or concentrated in nebular knots rather than appearing in the CS, which suggest that the WELS classification may be spurious.

In this paper, we present new results obtained with the first IFS study of the PN Hen~2-108. Our data allows for detailed study of the usual diagnostics, resulting in the first spatially resolved emission line flux, extinction coefficient, temperature and density maps of this object. The high quality integrated spectrum is used to study in detail the abundances, which are compared to the literature results mentioned previously. We also extracted the stellar spectrum and evaluate the different ambiguous classifications of the object, as well as its Wolf-Rayet like [WR] nature with detailed models. We combine the IFS spatially-resolved spectrum with pseudo-3D photoionization modelling to reveal new details of the Hen~2-108 density distribution.

This paper is organized as follows. We present the observation properties and the data reduction process in Section~\ref{sec:observations}. In Section~\ref{sec:spatialanalisys} the discussion of the flux, electronic density and temperature maps are presented. The integrated spectrum, the analysis of the emission line fluxes, the nebular emission and the physical and chemical parameters obtained from them are given in Section~\ref{sec:integratedspectrum}. We use popular tools extensively used by the community {\sc NEAT} \citep{Wesson12} and {\sc PyNeb} \citep{Luridiana15} to analyse the data, and we compare the with the results obtained by the {\sc Satellite} software \citep{Akras20,Akras22}, presented in Section~\ref{sec:satellite}, that evaluate an analysis of the spatial variation of physical parameters and chemical abundances. The spectrum of the central region and the analysis of the features observed in that region are given in Section~\ref{sec:CSspec}. We present a study of the PN parameters, ionization structure and matter distribution of Hen~2-108 using pseudo-3D photoionization models in Section~\ref{sec:structuremodels}. The general concluding discussion is given in Section~\ref{sec:conclusions}.

\section{Observations and data reduction}
\label{sec:observations}


The observations analysed here were obtained with the Visible Multi-Object Spectrograph \citep[VIMOS,][]{Lefevre03}, mounted on ESO-VLT UT3 Melipal, Paranal Observatory, in Chile, between Mayl 8th, 2007  and September 1st, 2007. The data are part of the observation program 079.D-0117A, led by H. Schwarz (\textit{in memoriam}). The observations were performed with airmasses of $\approx 1.6$ and $\approx 1.5$ for the blue and red arms, respectively. The average seeing obtained from the DIMM was of 0.89 and 0.69~arcsec for the blue and red arms respectively. Details for individual data files can be obtained from the public ESO archive at \url{http://archive.eso.org/eso/eso_archive_main.html}.

VIMOS offers two spatial samplings of 0.33~arcsec (2:1 magnification) and 0.67~arcsec (1:1 magnification) which we use in these observations. A shutter was used to mask the IFU and use only 40$\times$40 fibers. This provides us with a field of view (FoV) of 13$\times$13~arcsec for the 2:1 magnification and a large FoV of 27$\times$27~arcsec for the 1:1 magnification. Using the available high spectral resolution modes HRorange and HRblue, we obtained a wavelength coverage from 4000-6700 and 5250-7400\AA\ with a dispersion of 0.53 and 0.6~\AA~per pixel, respectively. The regions observed generate two separated data cubes at the end of the reduction process. The spectral resolution achieved is approximately 2500, with 4096 pixels in the spectral axis. The data obtained with the 2:1 magnification was only used to analyse the central source since it did not contain the entire nebula due to pointing problems. These characteristics and exposure information are shown in Table \ref{tab:obsdata}.

The data reduction was performed with the VIMOS Interactive Pipelines \citep[VIPGI;][]{Scodeggio05}, using standard procedures of bias and flat field correction, as well as wavelength and flux calibration. For the flux calibration, the standard star used was the white dwarf EG 274. The weather data from ESO showed that for the dates observed the sky was clear and for most of the night photometric. The data was also corrected for differential atmospheric refraction and sky contamination. For the first one, we used a Python script that, with a selected reference star in the observed field, traces its spatial position along each cut in the wavelength axis. The reference star position is obtained with a fit of a Gaussian function to the continuum map and identifying its maximum. The routine determines displacements of the star relative to a reference pixel of choice for all wavelengths. The coordinates of wavelength and displacements of the star are then fit by a second-order polynomial, which is used to correct the positions in each wavelength slice in the data cube.

The sky contamination was determined from a region in the data cube external to the ionized gas emission (see the yellow rectangle in Fig.~\ref{fig:flux108-2}). With the sky region selected, the median of the flux was obtained in each spaxel, resulting in the estimated sky spectrum per fiber. This sky spectrum was subtracted from all spaxels of the data cube. The residual sky in the data due to errors in our procedure is no larger than 9~percent at the [\ion{O}{i}]~6300~\AA\ line position. We have also inspected maps of the skylines and found no significant slope in the sky contribution, which justifies the use of a median value.

\begin{table*}
\caption{Table with details of the observational data used}
\label{tab:obsdata}
\resizebox{\textwidth}{!}{%
\begin{tabular}{llllllllll}
\hline
Grism     & Filter & Coverage   & Resolution & Dispersion & Seeing & Magnification & FOV     & Exposure & Date \\
          &        &      &            & \AA/pix      & DIMM   &               &         &          \\
\hline
HR blue   & Free   & 400-670 nm & 1440       & 0.53       & 0.89   & 1:1           & 27"x27" & 4x160s   & 2007-08-09 \\
HR orange & GG435  & 525-740 nm & 2650       & 0.60       & 0.69   & 1:1           & 27"x27" & 1x100s   & 2007-09-01 \\
HR blue   & Free   & 400-670 nm & 1440       & 0.53       & 0.89   & 2:1$^1$           & 13"x13" & 2x160s   & 2007-05-08 \\
HR orange & GG435  & 525-740 nm & 2650       & 0.60       & 0.69   & 2:1$^1$           & 13"x13" & 3x100s   & 2007-05-08 \\
\hline
\multicolumn{10}{c}{$^1$ Data used only for the central source analysis.}
\end{tabular}
}%
\end{table*}

\section{Spatially Resolved Analysis} \label{sec:spatialanalisys}

\subsection{Emission line flux maps}\label{sec:LineMaps}

\begin{figure*}
\centering
\includegraphics[width=\textwidth]{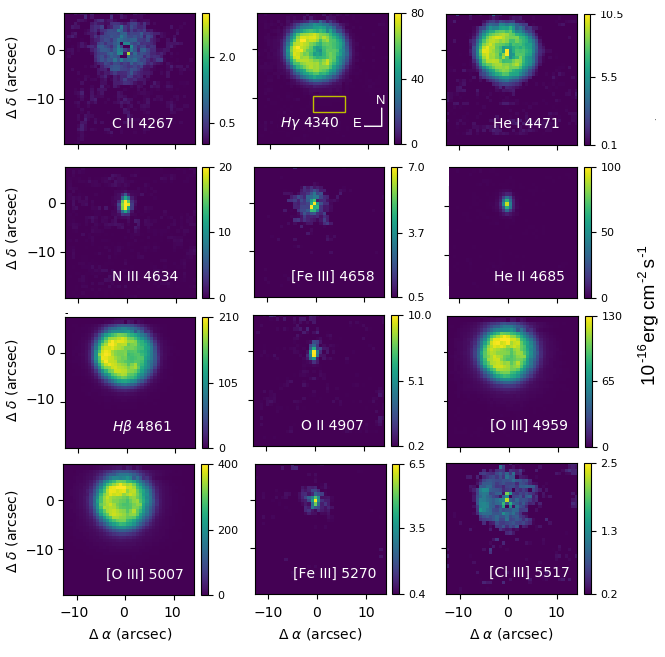}
\caption{Hen~2-108 flux maps extracted from the VIMOS/VLT observations. The flux is not corrected for extinction. The axes indicate the position offsets measured from the PN centre. All maps in this paper are orientated with north up and east to the left, as indicated in the H$\alpha$ map. The yellow rectangle in the H$\alpha$ map delimits the region used to obtain the integrated sky spectrum. The dark pixel to the north of central region in some red wavelengths is caused by a dead spaxel in the cube.}
\label{fig:flux108-2}
\end{figure*}
\begin{figure*}
\begin{center}
\includegraphics[width=\textwidth]{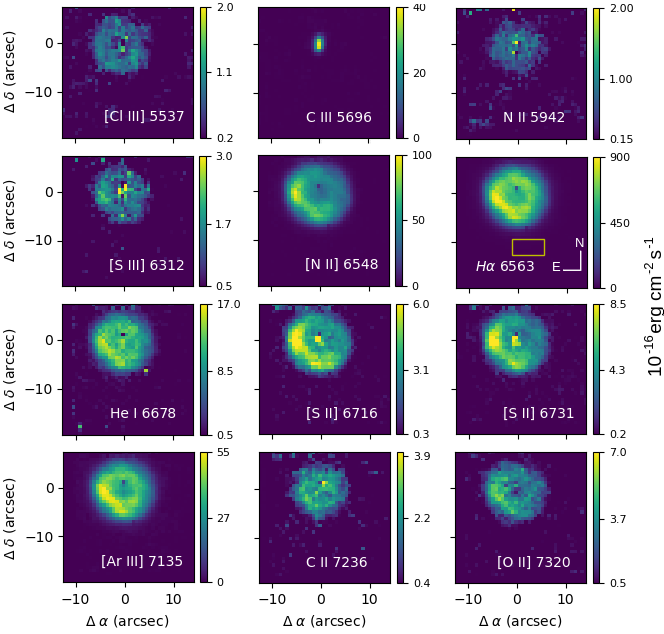}
\end{center}
\contcaption{}
\end{figure*}

The emission line flux maps obtained from the 1:1 magnification cube are shown in Fig.~\ref{fig:flux108-2}. To obtain the emission line maps we used a script that reads in the data cubes and performs Gaussian fits to selected  emission lines in each spaxel. For a given emission line, a fit is performed in a specified region of the spectra extracted around the line centre. The fits are done using the Levenberg-Marquardt least-squares algorithm. The script outputs maps of the emission line flux, central wavelength, width and continuum.

Analysing the morphology in the maps, two main structures are immediately identified: a central region and a surrounding ring structure. Some lines are produced in both structures, while others exclusively in one of them. 

The lines in Fig.~\ref{fig:flux108-2} \ion{N}{iii} $\lambda$4634, [\ion{Fe}{iii}] $\lambda$4658 and $\lambda$5270, \ion{C}{iii} $\lambda$5696, \ion{O}{ii} $\lambda$4907 and \ion{He}{ii} $\lambda$4686 are emitted predominantly in the central region of the nebula. This region has a FWHM of $2.7\pm0.3$~arcsec across the spectral range. For comparison, we also obtained the FWHM of the standard star used in our calibrations, obtaining a FWHM of $2.5\pm0.3$~arcsec. These values are consistent within uncertainties indicating that the emission is unresolved. It is interesting to highlight the difference between the \ion{N}{ii} $\lambda$5942 flux map and the \ion{N}{iii} $\lambda$4634 map, with extended emission in the first one and emission just in central region in the other one. 

Based on the emission line maps in Fig.~\ref{fig:flux108-2} we can see that the lines \ion{C}{ii} $\lambda$4267 and \ion{C}{ii} $\lambda$7236 have a extended morphology indicating the nebular origin. However the line \ion{C}{iii} $\lambda$5696 is concentrated on the central region. Interestingly, \citet{Gorny14} identified \ion{C}{ii} $\lambda$7236 as stellar.

The FWHM of the profile in the emission line map of \ion{He}{ii} $\lambda4686$ in Fig.~\ref{fig:flux108-2} is $\sim$2.3~arcsec. The presence of this \ion{He}{ii} line is unexpected, as CSs with effective temperatures around 40~kK cannot produce a large zone of \ion{He}{ii} and, therefore, significant emission from this ion. To produce the observed \ion{He}{ii} $\lambda4686$/H$\beta$ ratio, a star with a temperature of at least $\sim$60~kK is required \citep[e.g.][see also the discussion in Section~\ref{sec:structuremodels}]{Gurzadian1988}.

The line of [\ion{Cl}{iii}] $\lambda5517$ has a low flux in comparison to the others represented in Fig.~\ref{fig:flux108-2}. In addition to the emission in the central region, there is fainter extended emission around. The eastern side of the nebula emits a slightly larger flux than the western side.

The H$\alpha$, H$\beta$, H$\gamma$, [\ion{N}{ii}], [\ion{Ar}{iii}], [\ion{O}{iii}] and [\ion{O}{ii}] in Fig.~\ref{fig:flux108-2} show an annular structure, i.e., less flux in the central region of the nebula and greater brightness in the surrounding region. In all maps, we can see that the flux is not uniform, this may indicate that the matter is not homogeneously distributed in this ring. In almost all maps, except the [\ion{O}{iii}], we can distinguish a brighter region in the southeast region of the ring. For [\ion{O}{iii}], a bright spot is seen in the northern region.

For [\ion{S}{ii}] and \ion{He}{i}, the emission is produced both in the central and the ring structures. In \ion{He}{i} $\lambda6678$, the central region is slightly distinguished.

\subsection{Extinction map} \label{sec:extinction}

An important step in analysing the data is the correction for interstellar extinction. Since we have spatially resolved data, we can obtain spatial information on this parameter. Adopting the extinction law of \citet{Cardelli89}, Rv=3.1, and using the observed ratio map of H$\alpha$/H$\beta$, we obtained the extinction map for Hen~2-108 with {\sc PyNeb}  \citep{Luridiana15} considering the \cite{Chianti21} atomic data. The extinction is obtained by comparing the observed ratio of H$\alpha$/H$\beta$ to its theoretical value of 2.85 taken from \citet{Osterbrock06} for temperature and density values adequate to the object being studied. The temperature and density are obtained simultaneously from an initial guess by fitting  [\ion{S}{ii}] $\lambda$6731/$\lambda$6716 and [\ion{N}{ii}] $\lambda$5755/$\lambda$6548 \citep{Luridiana15}. We performed one iteration with the estimated temperature and density to converge on the correct Ha/Hb ratio and then recalculate the physical parameters as recommended by \citet{Ueta21}.

\begin{figure}
\includegraphics[width=\linewidth]{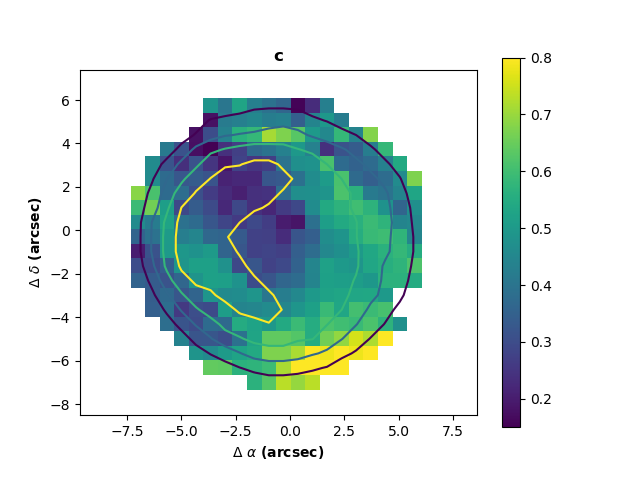}
\caption{Hen~2-108 extinction coefficient map obtained from H$\alpha$/H$\beta$ ratio. The contour lines represent the isocontours at 30, 50, 70 and 90 per cent of the H$\beta$ flux maximum value on the map. The axes indicate the position offsets from the PNe centre in arcsec.}
\label{fig:extinction}
\end{figure}

In order to avoid unrealistic ratio values due to low signal-to-noise data in the outer parts of the nebula, we have limited the observed region used in the spatially resolved diagnostic analysis. The nebular emission region used for the analysis was determined such that the H$\beta$ emission had a spaxel signal-to-noise of at least 3. To achieve this, we have applied a mask to the emission line maps such that spaxels that did not satisfy the criteria were not considered. This mask was applied in all diagnostic maps.
The resulting $c$(H$\beta$) map is shown in Fig.~\ref{fig:extinction}. The H$\beta$ flux contour lines are overplotted to help find possible relations between the extinction and the nebular structures. In the nebular region, there is a variation in the value of $c$(H$\beta$) between 0.3 and 0.6. Differently from what is seen in the H$\beta$ and other line maps, it is not possible to clearly distinguish the annular structure in the extinction map.

\subsection{Electronic density and temperature maps}

After correcting for extinction using the $c$(H$\beta$) map in Fig.~\ref{fig:extinction}, we use the line flux maps of [\ion{S}{ii}] $\lambda \lambda6716,6731$ and [\ion{N}{ii}] $\lambda \lambda5755,6584$ to calculate the maps of the electronic density ($N_e$) and temperature ($T_e$) using the {\sc PyNeb} $diags.getCrossTemDen$ tool. The Hen~2\nobreakdash-108 electronic temperature map was calculated from the [\ion{N}{ii}] $\lambda\lambda$5755/6584 line ratio iteratively with the density map obtained from the [\ion{S}{ii}] $\lambda\lambda$6716/6731 line ratio. Based on the model results presented in Section~\ref{sec:structuremodels}, we find that the known contribution of recombination of N$^{++}$ is under 1\% to the $\lambda$5755 line intensity, so the recombination line contribution will be disregarded.

\begin{figure}
\includegraphics[width=\linewidth]{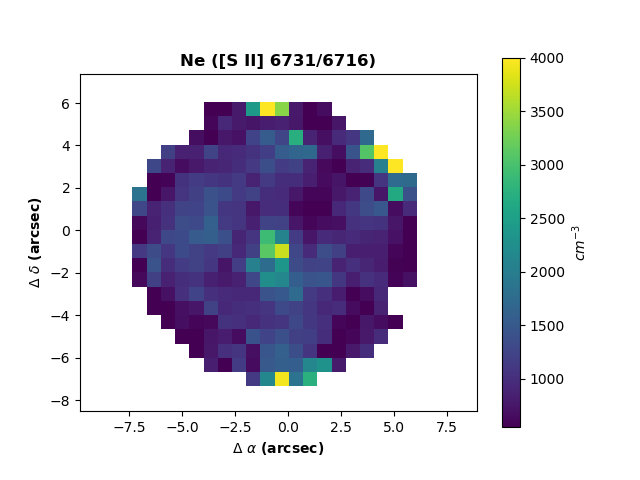}
\includegraphics[width=\linewidth]{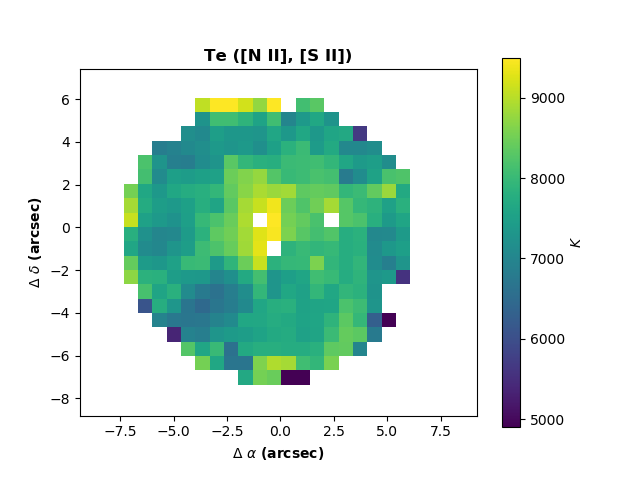}
\caption{Hen~2-108 density (in cm$^{-3}$; upper panel) and temperature (in K; bottom panel) maps derived using the [\ion{S}{ii}]$\lambda\lambda$6716,6731 and [\ion{N}{ii}]$\lambda\lambda$5755,6584 emission lines, respectively. The axes indicate the position offsets from the PNe centre in arcsec.} \label{fig:denstemp-He2-108}
\end{figure}

Figure~\ref{fig:denstemp-He2-108} shows the resulting Hen~2-108 density and temperature spatial distribution. Most $N_e$ values are between $\sim$800 and $\sim$2000 cm$^{-3}$, except for the central region, where the density reaches values up to $\sim$3000~cm$^{-3}$. The central region also shows higher temperatures than the rest of the nebula, with a difference of over 1000~K. The values for $N_e$ and $T_e$ in these central regions should be taken with care as there may be some effect from the noise of the continuum of the central source affecting the line intensities in those pixels. Some spaxels with extreme values close to the border of the useful region still remain, but they are likely the result of low signal of the lines used in the diagnostics which are not exactly coincident with H$\beta$,  used to define the signal-to-noise cut-off.


\section{Analysis of the integrated spectrum}
\label{sec:integratedspectrum}

\begin{figure*}
  \centering
  \includegraphics[width=16cm,trim={0cm 0.4cm 0 0.5cm},clip]{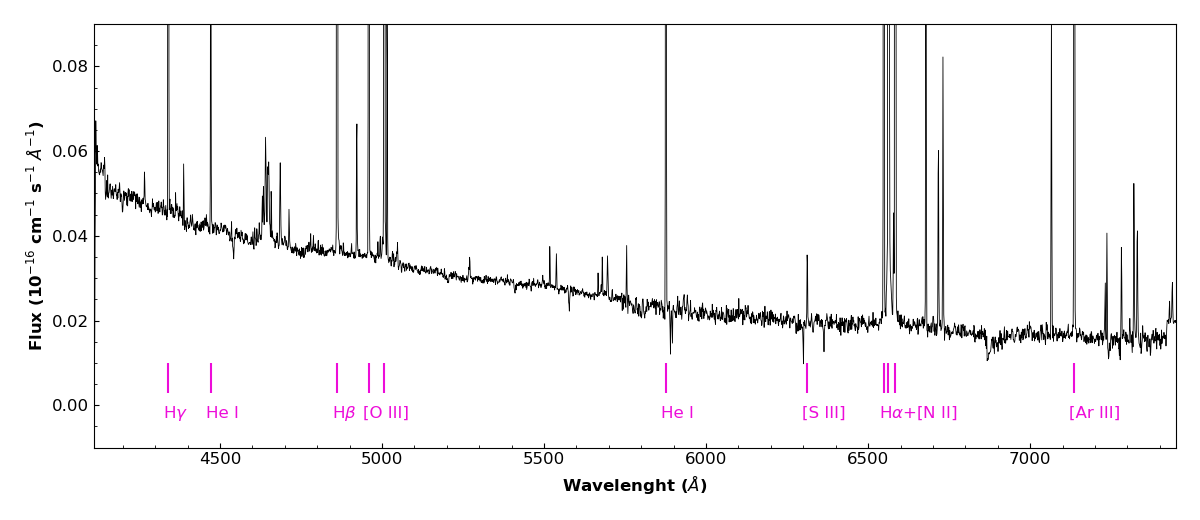}
  \caption{Hen~2-108 integrated spectrum. Some of the nebular emission lines are identified.}\label{fig:integratedspectrum}
\end{figure*}

The integrated spectrum of Hen~2-108 was obtained from the data cube with magnification 1:1 by summing over the spaxels in the region delimited by the significant H$\beta$ mask as described in Section~\ref{sec:extinction}. The resulting spectrum is shown in Fig.~\ref{fig:integratedspectrum}, where some of the Hen~2-108 emission lines are identified.

The observed set of ions seen in the spectrum are that typically seen in a low-ionization nebula. We identified 53 emission lines from 10 elements (15 ions). The brightest lines (above 20\% of H$\beta$) found in the spectrum are \ion{H}{i} and \ion{He}{i} recombination lines and [\ion{O}{iii}], [\ion{N}{ii}], and [\ion{Ar}{iii}] forbidden lines. 

\begin{figure*}
  \centering
  \includegraphics[scale=0.51,trim={0cm 0.6cm 0 0.2cm},clip]{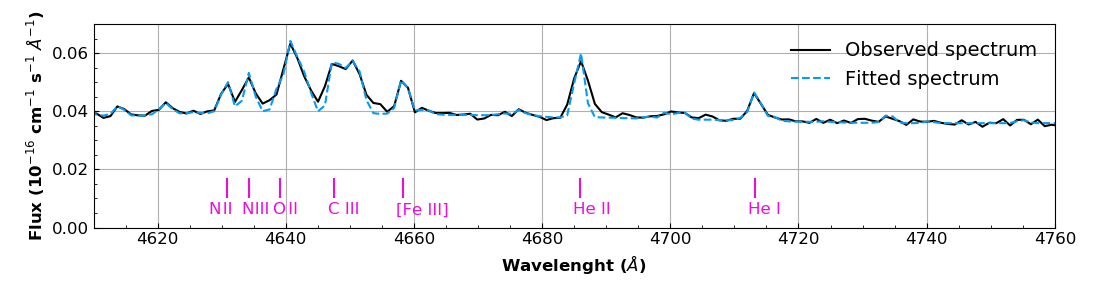}
  \includegraphics[scale=0.51,trim={0cm 0.6cm 0 0.2cm},clip]{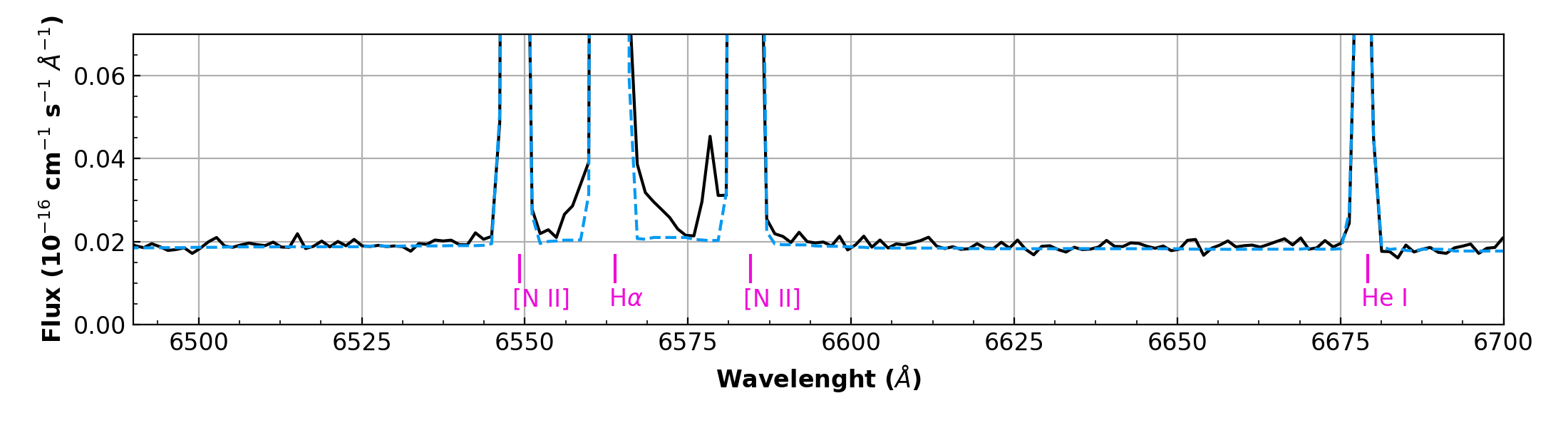}
  \caption{Sections of the Hen~2-108 observed integrated spectrum (black solid curve) and the corresponding ALFA fitting (cyan dashed curve). The carriers of emission lines are indicated.} \label{qualialfa1}
\end{figure*}

We also highlight the presence of a cluster of weak lines between 4630 and 4660~\AA\ (hereafter called the $\lambda$4630 complex), which is shown in details in Fig.~\ref{qualialfa1}. The $\lambda$4630 complex is composed mostly of metal recombination lines produced in the central region of Hen~2-108. These lines were identified as \ion{N}{ii} $\lambda4630$, \ion{N}{iii} $\lambda4634$, \ion{O}{ii} $\lambda4639$, blend of \ion{C}{iii}, \ion{O}{ii} and \ion{C}{iii} $\lambda4647,4649,4650$ and [\ion{Fe}{iii}] $\lambda4658$. The maps of two of these lines can be seen in Fig.~\ref{fig:flux108-2}.

The comparison between the H$\alpha$ observed line profile and the Gaussian fitting, displayed in details in Fig.~\ref{qualialfa1}, shows indication that this line has wide wings. We found no clear evidence of wide wings on other lines. Some other interesting features are discussed in more detail in Section~\ref{sec:CSspec}, where we analyse the spectrum of the PN central region.

\subsection{Emission line fluxes}
 
To measure the line fluxes, we used the code {\sc ALFA} \citep[Automated Line Fitting Algorithm, version 1.0.140;][]{Wesson16}. The code also suggests line identifications, which were confirmed against other publications in the literature \citep[e.g.,][]{Osterbrock06, Van_Hoof_2018, Aleman_etal_2019_Tc1}. The {\sc ALFA} fitting of the Hen~2-108 spectrum was made with no normalization and the following parameter values: rebin of 2; 2000 generations; population size of 2000; pressure of 0.1 and size of continuum window of 51 pixels. We adopted these parameter values to improve signal-to-noise and the detection performance of {\sc ALFA}. The overall quality level of the {\sc ALFA} fitting is exemplified in Fig.~\ref{qualialfa1}, where part of the spectrum is shown in more details. The {\sc ALFA} fitting provides a relative residual typically less than 15 per cent for most of the spectrum\footnote{Higher values occur in regions of the continuum, due to its low flux, and in the wings of some lines, due to deviation of the Gaussian profile. Although the residual may reach up to 40 percent in the line wings, it does not influence the line fluxes, as the flux in the wings do not contribute significantly to the total line flux in our case. High values were also observed where the sky contribution was not perfectly removed, but also in these cases, they do not influence significantly the line fluxes we report.}. 

Table~\ref{tab:fluxlinesNP} lists the measured line fluxes and the uncertainties based on the {\sc ALFA} fitting procedure. We only include lines that are detected with signal-to-noise ratio S/N$>$3. The S/N ratio is obtained from the residuals determined by {\sc ALFA} from the difference of the fitted spectrum with respect to the observed spectrum. The observed and rest wavelengths, as well as the line carrier, are shown in the first three columns. The fourth and fifth columns present, respectively, the measured fluxes and their errors, while the sixth and seventh columns present the corresponding values but corrected from reddening. 
Columns eight to ten show measured fluxes available in the literature for comparison. With a few exceptions that are discussed in Section~\ref{sec:module_specific_slits}, our integrated fluxes are very similar to the values found in the literature. All fluxes are scaled to H$\beta=$~100. The Hen~2-108 absolute H$\beta$ flux derived from our integrated spectrum is $(1.9\pm 0.3) \times 10^{-12}$~erg~cm$^{-2}$~s$^{-1}$ (not corrected for extinction) and $(7.0\pm 1.1) \times 10^{-12}$~erg~cm$^{-2}$~s$^{-1}$ (corrected for extinction).

From the fitting procedure, {\sc ALFA} also estimates the radial velocity of the object using all fitted lines. For Hen~2-108, the estimated barycentric radial velocity, corrected for the Earth's orbital and rotational velocities at the time of the observation, is -10.8~km~s$^{-1}$. Since {\sc ALFA} does not provide an uncertainty, we roughly estimated one of 3~km~s$^{-1}$ based on the spectral sampling.

\begin{table*}
\caption{Optical emission line fluxes (scaled to H$\beta=$~100.).} \label{tab:fluxlinesNP}

\begin{tabular}{ccccccccccc}

\hline

$\lambda_\textrm{obs}$ & $\lambda_\textrm{rest}$ &	Ion	& $F(\lambda)$ & $\sigma_{F(\lambda)}$ & $I(\lambda)$ & $\sigma_{I(\lambda)}$ & A90	& G14 & {\sc Satellite} & {\sc Satellite} \\
(\AA) & (\AA)&		&   &   &  &  &  &  & (G14's slit) & (integrated) \\
\hline

4267.44	&	4267.15	&	 C~{\sc ii}       	&	0.47	&	   0.09	&	0.58	&	 0.11  	&	--	&	--	&	--	&	--\\
4340.77	&	4340.47	&	 H~{\sc i}        	&	35.52	&	   0.18	&	43.01	&	 0.23   &	44	&	46.88 &	--	&	--	\\
4363.51	&	4363.21	&	 [O~{\sc iii}]    	&	0.17	&	   0.05	&	0.20	&	 0.06 	&	--	&	1.3 $^{a}$ &	--	&	--	\\
4388.23	&	4387.93	&	 He~{\sc i}       	&	0.68	&	   0.06	&	0.80	&	 0.07 	&	--	&	-- &	--	&	--	\\
4409.60	&	4409.30	&	 Ne~{\sc ii}      	&	0.10	&	   0.02	&	0.12	&	 0.02 	&	--	&	-- &	--	&	--	\\
4415.20	&	4414.90	&	 O~{\sc ii}       	&	0.17	&	   0.02	&	0.20	&	 0.02 	&	--	&	-- &	--	&	--	\\
4471.66	&	4471.50	&	 He~{\sc i}       	&	4.68	&	   0.09	&	5.35	&	 0.10 	&	--	&	6.27 $^{b}$	&	--	&	--\\
4515.02	&	4514.86	&	 N~{\sc iii}      	&	0.14	&	   0.04	&	0.16	&	 0.04 	&	--	&	--	&	--	&	--\\
4607.20	&	4607.03	&	 [Fe~{\sc iii}]   	&	0.12	&	   0.04	&	0.13	&	 0.05 	&	--	&	--	&	--	&	--\\
4621.56	&	4621.39	&	 N~{\sc ii}       	&	0.18	&	   0.03	&	0.17	&	 0.03 	&	--	&	--	&	--	&	--\\
4630.71	&	4630.54	&	 N~{\sc ii}       	&	0.53	&	   0.09	&	0.52	&	 0.10 	&	--	&	--	&	--	&	--\\
4634.31	&	4634.14	&	 N~{\sc iii}      	&	0.84	&	   0.11	&	0.63	&	 0.12 	&	--	&	--	&	--	&	--\\
4639.03	&	4638.86	&	 O~{\sc ii}       	&	0.43	&	   0.12	&	0.46	&	 0.13 	&	--	&	--	&	--	&	--\\
4647.59	&	4647.42	&	 C~{\sc iii}      	&	0.69	&	   0.13	&	0.74	&	 0.14 	&	--	&	--	&	--	&	--\\
4649.30	&	4649.13	&	 O~{\sc ii}       	&	0.54	&	   0.11	&	0.58	&	 0.12 	&	--	&	--	&	--	&	--\\
4650.42	&	4650.25	&	 C~{\sc iii}      	&	0.43	&	   0.10	&	0.56	&	 0.12 	&	--	&	--	&	--	&	--\\
4651.64	&	4651.47	&	 C~{\sc iii}      	&	0.33	&	   0.10	&	0.48	&	 0.11 	&	--	&	--	&	--	&	--\\
4658.27	&	4658.10	&	 [Fe~{\sc iii}]   	&	0.53	&	   0.05	&	0.57	&	 0.05 	&	--	&	--	&	--	&	--\\
4685.85	&	4685.68	&	 He~{\sc ii}      	&	1.14	&	   0.11	&	1.06	&	 0.12 	&	--	&	3.73 &	4.67	&	1.92\\
4788.31	&	4788.13	&	 N~{\sc ii}       	&	0.17	&	   0.04	&	0.17	&	 0.04 	&	--	&	--	&	--	&	--\\
4861.51	&	4861.33	&	 H~{\sc i}        	&	100  	&	   0.14	&	100	    &	 1.00  	&	100	&	100	&	100	&	100\\
4881.34	&	4881.11	&	 [Fe~{\sc iii}]   	&	0.13	&	   0.02	&	0.13	&	 0.02 	&	--	&	--	&	--	&	--\\
4922.17	&	4921.93	&	 He~{\sc i}       	&	1.65	&	   0.04	&	1.62	&    0.04 	&	--	&	--	&	--	&	--\\
4959.15	&	4958.91	&	 [O~{\sc iii}]    	&	56.57	&	   0.22	&	55.90	&	 0.10  	&	--	&	58.95 &	58.6	& 57.5	\\
5007.08	&	5006.84	&	 [O~{\sc iii}]    	&	175.55	&	   0.35	&	168.00 	&	 1.00  	&	167	&	176.81 & 176 &	173	\\
5015.92	&	5015.68	&	 He~{\sc i}       	&	3.39	&	   0.21	&	3.24	&    0.20 	&	--	&	--	&	--	&	--\\
5047.98	&	5047.74	&	 He~{\sc i}       	&	0.35	&	   0.05	&	0.33	&    0.05 	&	--	&	--	&	--	&	--\\
5270.65	&	5270.40	&	 [Fe~{\sc iii}]   	&	0.26	&	   0.04	&	0.23	&	 0.04 	&	--	&	--	&	--	&	--\\
5517.92	&	5517.66	&	 [Cl~{\sc iii}]   	&	0.54	&	   0.03	&	0.46	&    0.03 	&   --	&	0.43 $^{b}$	&	0.44	&	0.49\\
5537.86	&	5537.60	&	 [Cl~{\sc iii}]   	&	0.48	&	   0.02	&	0.40	&	 0.02 	&	--	&	0.34 $^{b}$	&	0.33	&	0.46\\
5666.90	&	5666.63	&	 N~{\sc ii}       	&	0.25	&	   0.03	&	0.20	&	 0.03 	&	--	&	--	&	--	&	--\\
5679.85	&	5679.56	&	 N~{\sc ii}       	&	0.44	&	   0.06	&	0.36	&	 0.05 	&	--	&	--	&	--	&	--\\
5696.21	&	5695.92	&	 C~{\sc iii}      	&	0.51	&	   0.06	&	0.45	&	 0.05 	&	--	&	--	&	--	&	--\\
5711.04	&	5710.77	&	 N~{\sc ii}       	&	0.16	&	   0.03	&	0.13	&	 0.02 	&	--	&	--	&	--	&	--\\
5754.90	&	5754.60	&	 [N~{\sc ii}]     	&	0.77	&	   0.02	&	0.62	&	 0.02 	&	--	&	0.56 &	0.62 &	0.67\\
5875.96	&	5875.66	&	 He~{\sc i}       	&	20.87	&	   0.12	&	{16.46}	&	 {0.10} 	&	17.5&	16.77	&	16.80	&	16.81\\
5932.08	&	5931.78	&	 N~{\sc ii}       	&	0.25	&	   0.06	&	{0.20}	&	 0.05 	&	--	&	--	&	--	&	--\\
5941.95	&	5941.65	&	 N~{\sc ii}       	&	0.27	&	   0.07	&	0.21	&	 0.05 	&	--	&	--	&	--	&	--\\
6312.37	&	6312.10	&	 [S~{\sc iii}]    	&	0.92	&	   0.07	&	{0.68}	&	 0.05 	&	--	&	0.79 &	0.93	& 0.76	\\
6548.38	&	6548.10	&	 [N~{\sc ii}]     	&	36.69	&	   0.72	&	{25.96}	&	 {0.51} 	&	--	&	21.98	&	25.09	&	26.81 \\
6563.05	&	6562.77	&	 H~{\sc i}        	&	417.16	&	   0.75	&	{295.00} 	&	 1.00  	&	310	&	285.61	&	296 &	297\\
6583.78	&	6583.50	&	 [N~{\sc ii}]     	&	112.97	&	   1.88	&	{79.50}	&	 {1.30} 	&	90	&	69.73	&	77.71	&	82.10\\
6678.47	&	6678.16	&	 He~{\sc i}       	&	6.84	&	   0.07	&	{4.74}	&	 {0.05} 	&	--	&	4.85 &	5.21	& 5.02	\\
6716.75	&	6716.44	&	 [S~{\sc ii}]     	&	2.66	&	   0.06	&	{1.83}	&	 0.04 	&	--	&	2.01	&	2.12	&	2.16\\
6731.13	&	6730.82	&	 [S~{\sc ii}]     	&	3.49	&	   0.08	&	{2.40}	&	 0.05 	&	--	&	2.41	&	2.58	&	2.64\\
7065.58	&	7065.25	&	 He~{\sc i}       	&	4.65	&	   0.06	&	{3.04}	&	 0.04 	&	--	&	--	&	--	&	--\\
7136.12	&	7135.80	&	 [Ar~{\sc iii}]   	&	24.32	&	   0.11	&	{15.71}	&	 {0.07} 	&	15	&	15.39	&	16.20	&	16.31\\
7231.65	&	7231.32	&	 C~{\sc ii}       	&	0.72	&	   0.11	&	{0.46}	&	 0.07 	&	--	&	--	&	--	&	--\\
7236.52	&	7236.19	&	 C~{\sc ii}       	&	1.17	&	   0.11	&	{0.74}	&	 {0.07} 	&	--	&	--	&	--	&	--\\
7281.68	&	7281.35	&	 He~{\sc i}       	&	1.34	&	   0.08	&	0.85	&	 0.05 	&	--	&	--	&	--	&	--\\
7320.32	&	7319.99	&	 [O~{\sc ii}]     	&	2.28	&	   0.10	&	{1.43}	&	 0.06 	&	--	&	2.78 $^{c}$	& 1.46	& 1.50\\
7330.00	&	7330.20	&	 [O~{\sc ii}]     	&	0.99	&	   0.08	&	{0.62}	&	 0.05 	&	--	&	--	&	1.30	& 1.42\\
7331.06 &   7330.73 &	 [O~{\sc ii}]     	&   1.02  	&	   0.08	&  	{0.64}	&	 0.05 	&	--	&	--	&	 	&  \\	

\hline

\multicolumn{11}{c}{Notes: $^{(:)}$ Uncertain values. $^{(a)}$ Line flux error at 40\%. $^{(b)}$ Line flux error at 20\%. $^{(c)}$ Total flux of the doublet}\\

\multicolumn{11}{c}{[O~{\sc ii}]~$\lambda\lambda$7320+7330 lines. References: A90: private communication with A. Acker published in}\\

\multicolumn{11}{c}{\citet{Pottasch11} (P11). G14: \citet{Gorny14}.}

\end{tabular}
\end{table*}


\begin{table}
\centering
\caption{Main diagnostic results for Hen~2-108 compared with literature data.}
\label{tab:diagNP108}

\resizebox{\columnwidth}{!}{%

\begin{tabular}{lcccc}
\hline 
    &	 \multicolumn{3}{c}{This work}	                    & G14 \\
\hline 
	&	{\sc NEAT}	&	{\sc Satellite}	&	{\sc Satellite} & \\
	&		        &	integrated	    &	G14's slit      & \\
	
\hline  													

\multicolumn{5}{l}{\textbf{Extinction:}}\\

c(H$\beta)$                      	&	  {0.508 }  $\pm$ 0.001 	&	 0.49 $\pm$ 0.05  	&	 0.44  $\pm$ 0.06  		&	0.51	\\

\\
	
\multicolumn{5}{l}{\textbf{Densities:}}\\

$[$\ion{S}{ii}$]$ $\frac{\lambda6731}{\lambda6717}$       &	  1.31 $\pm$ 0.04 &	1.22 $\pm$ 0.08 & 1.21 $\pm$ 0.11	 	&	1.20	\\
$n_\textrm{e}$[\ion{S}{ii}] (cm$^{-3}$)   	&	  {1540} $\pm$ {170}   	&	 1350 $\pm$ 120 	&	 1319 $\pm$ 150   	&	871	\\
\\
$[$\ion{Cl}{iii}] $\frac{\lambda5537}{\lambda5517}$          	&	  {0.89}   $\pm$ 0.06 	& 0.94 $\pm$ 0.12		& 0.73 $\pm$ 0.13		&	0.79:	\\
$n_\textrm{e}$[\ion{Cl}{iii}] (cm$^{-3}$) 	&	{992 $\pm$ 361} 	&	  1594  $\pm$ 630   	& 325 $\pm$ 140  	 	&	--   	\\
\\
\multicolumn{5}{l}{\textbf{Temperature:}}	\\
$[$\ion{N}{ii}] $\frac{(\lambda6548+\lambda6584)}{\lambda5754}$ 	&	 {170  $\pm$ 5} 	& 185 $\pm$ 4		& 183 $\pm$ 12			  	&	164	\\
$T_\textrm{e}$[\ion{N}{ii}]  (K)      	&	{7890} $\pm$ {70} 	&	 8027 $\pm$ 110  	&	 8034 $\pm$ 160  		&		10396	\\
\\
$[$\ion{O}{iii}] $\frac{(\lambda4959+\lambda5007)}{\lambda4363}$ 	&	 {1100  $\pm$ 440} 	& - $\pm$ -		& - $\pm$ -			  	&	-	\\
$T_\textrm{e}$[\ion{O}{iii}]  (K)      	&	{6610} $\pm$ {410} 	&	 - $\pm$ -  	&	 - $\pm$ -  		&		-	\\
\hline
\multicolumn{5}{c}{References: G14: \citet{Gorny14}. Note: (:) Uncertain value.}\\
\end{tabular}%
}
\end{table}


\begin{table*}
\centering
\caption{Chemical abundances of Hen~2-108.}
\label{tab:abundNPs}
\begin{tabular}{lccccccccc}

\hline

Code	&	{\sc Neat}	&	\multicolumn{4}{c}{\sc Satellite}	&	P11 	&	 G14   	\\

Aperture$^{(*)}$	&	Integrated	&	\multicolumn{2}{c}{G14's Slit}	&	\multicolumn{2}{c}{Integrated} &	mixed 	&	Slit	\\

ICF Scheme$^{(**)}$	&	DI14	&	KB94	&	DI14	&	KB94	&	DI14	&	P11 	&	KB94+L00 	\\

\hline															
\multicolumn{8}{l}{\textbf{Abundances from Recombination Lines}}\\

He$^+$/H$^+$ ($\times10$)	&	 1.09 $\pm$ 0.01 	&	 \multicolumn{2}{c}{1.15 $\pm$ 0.08} 		&	 \multicolumn{2}{c}{1.14 $\pm$ 0.02}		&		&		\\

He$^{++}$/H$^+$	($\times10^{3}$) &	 0.91 $\pm$ 0.09 	&	 \multicolumn{2}{c}{3.69 $\pm$ 0.08}	&	 \multicolumn{2}{c}{1.52 $\pm$ 0.05}		&		&		\\

He/H ($\times10$)	&	 1.10 $\pm$ 0.01	&	 \multicolumn{2}{c}{1.18 $\pm$ 0.08}		&	 \multicolumn{2}{c}{1.15 $\pm$ 0.02} 	&	 1.1	&	 1.26$:$ 	\\
\\

C$^{++}$/H$^+$ (RL) ($\times10^{4}$) &	 5.09 $\pm$ 0.10 	&	 -- 	&	 -- 	&	 -- 	&	 -- 	&		&		\\
C$^{+++}$/H$^+$ (RL) ($\times10^{4}$) &	 4.53 $\pm$ 0.86 	&	 -- 	&	 -- 	&	 -- 	&	 -- 	&		&		\\

ICF(C) (RL)	&	  1.00 $\pm$ 0.00 	&	 -- 	&	 -- 	&	 -- 	&	 -- 	&		&		\\
C/H	($\times10^{4}$) &	 9.62 $\pm$ 1.35	&	 -- 	&	 -- 	&	 -- 	&	 -- 	&	 -- 	&	 -- 	\\

\\

N$^{++}$/H$^+$ (V3) ($\times10^{3}$) &	 1.81 $\pm$ 0.17 	&	 -- 	&	 -- 	&	 -- 	&	 -- 	&		&		\\
N$^{++}$/H$^+$ (V20) ($\times10^{3}$) &	 5.00 $\pm$ 1.00 	&	 -- 	&	 -- 	&	 -- 	&	 -- 	&		&		\\
N$^{++}$/H$^+$ (V28) ($\times10^{3}$) &	 2.54 $\pm$ 0.47 	&	 -- 	&	 -- 	&	 -- 	&	 -- 	&		&		\\
N$^{++}$/H$^+$ (RL) ($\times10^{3}$) &	 2.46 $\pm$ 0.29 	&	 -- 	&	 -- 	&	 -- 	&	 -- 	&		&		\\


ICF(N)	&	  1.36 $\pm$ 0.11 	&	 -- 	&	 -- 	&	 -- 	&	 -- 	&		&		\\
N/H	($\times10^{3}$) &	 3.34 $\pm$ 0.49	&	 -- 	&	 -- 	&	 -- 	&	 -- 	&	 -- 	&	 -- 	\\

\\
O$^{++}$/H$^+$ (RL) ($\times10^{3}$)  &     2.86  $\pm$ 0.36	&	 -- 	&	 -- 	&	 -- 	&	 -- 	&		&		\\
ICF(O) (RL)           &     1.53    $\pm$ 0.17	&	 -- 	&	 -- 	&	 -- 	&	 -- 	&		&		\\
O/H (RL) ($\times10^{3}$) &   4.38   $\pm$  0.75 	&	 -- 	&	 -- 	&	 -- 	&	 -- 	&	--	&	--	\\

\hline
\multicolumn{8}{l}{\textbf{Abundances from Collisionally Excited Lines}}\\

N$^+$/H$^+$	($\times10^{5}$) &	 3.25 $\pm$ 0.12	&	 3.01 $\pm$ 0.43 	&	 3.01 $\pm$ 0.43 	&	 3.12 $\pm$ 0.18 	&	 3.12 $\pm$ 0.18 	&		&\\

ICF(N)	&	  3.81 $\pm$ 1.12 	&	  1.96 $\pm$ 0.41 	&	  2.14 $\pm$ 0.24 	&	  1.93 $\pm$ 0.18 	&	  2.25 $\pm$ 0.21 	&		&		\\

N/H	($\times10^{4}$) &	 1.23 $\pm$ 0.36 	&	 0.59 $\pm$ 0.10 	&	 0.64 $\pm$ 0.03	&	 0.60 $\pm$ 0.03	&	 0.70 $\pm$ 0.03 	&	 0.60 	&	0.28\\
\\

O$^+$/H$^+$	($\times10^{4}$) &	 1.59   $\pm$ 0.14 	&	 1.51  $\pm$ 0.44 	&	 1.51  $\pm$ 0.44 	&	 1.54   $\pm$ 0.13 	&	 1.54   $\pm$ 0.13 	&		&\\

O$^{++}$/H$^+$	($\times10^{4}$) &	 3.03   $\pm$ 1.16	&	 1.41   $\pm$ 0.11 &	 1.41   $\pm$ 0.11	&	 1.38   $\pm$ 0.10	&	 1.38   $\pm$ 0.10 	&		&\\

ICF(O)	&	1.00   $\pm$ 0.01 	&	1.02   $\pm$ 0.21 	&	1.02   $\pm$ 0.21 	&	1.01   $\pm$ 0.18 	&	1.01   $\pm$ 0.18 	&		&		\\

O/H	($\times10^{4}$) &	 4.65   $\pm$ 1.18	&	 2.98   $\pm$ 0.49	&	 2.97  $\pm$ 0.48	&	 2.94   $\pm$ 0.18	&	 2.94   $\pm$ 0.18 	&	  2.8 	&	 1.91 \\

\\

S$^+$/H$^+$	($\times10^{7}$) &	 2.23    $\pm$ 0.09	&	 2.66    $\pm$ 0.38	&	 2.66    $\pm$ 0.38 	&	 2.67    $\pm$ 0.08	&	 2.67    $\pm$ 0.08 	&		&		\\

S$^{++}$/H$^+$ ($\times10^{6}$) & -- & 5.95 $\pm$ 1.45 	&	 5.95    $\pm$ 1.45 	&	 4.81    $\pm$ 0.35	&	 4.81    $\pm$ 0.35  	&		&		\\

ICF(S)	&	 5.92    $\pm$ 1.53 	&	 1.04    $\pm$ 0.35 	&	 1.01    $\pm$ 0.33 	&	 1.04    $\pm$ 0.11 	&	 1.01    $\pm$ 0.11 	&		&		\\

S/H	($\times10^{6}$) &	 1.32 $\pm$ 0.35 &	 6.47 $\pm$ 1.23 &	 6.26    $\pm$ 1.16 &	 5.28    $\pm$ 0.33 &	 5.12    $\pm$ 0.30 &	 8.1 	&	1.67${:}$  	\\

\\

Ar$^{++}$/H$^+$	($\times10^{6}$) &	4.50 $\pm$ 1.15	&	 2.38 $\pm$ 0.36	&	 2.38 $\pm$ 0.36	&	 2.41 $\pm$ 0.11	&	 2.41 $\pm$ 0.11 & & \\

ICF(Ar)	&	 1.12   $\pm$ 0.06 	&	 1.87       	&	 1.08   $\pm$ 0.21 	&	 1.87      	&	 1.06   $\pm$ 0.06 	&		&		\\

Ar/H ($\times10^{6}$) &	 5.04   $\pm$ 1.61 & 4.45   $\pm$ 0.55 &	 2.57   $\pm$ 0.28 &	 4.51   $\pm$ 0.15 &	 2.55   $\pm$ 0.10 &	 5.1 	&	 1.92${:}$	\\

\\

Cl$^{++}$/H$^+$	($\times10^{7}$) &	 2.23 $\pm$ 0.74 	&	 0.96 $\pm$ 0.10 	&	 0.96 $\pm$ 0.10	&	 1.28 $\pm$ 0.09 	&	 1.28 $\pm$ 0.09 	&	&	\\

ICF(Cl)	&	1.34 $\pm$ 0.04 	&	 --  	&	  1.32 $\pm$ 0.15 	&	 --  	&	  1.31 $\pm$ 0.11 &	 & \\

Cl/H ($\times10^{7}$) &	 2.97 $\pm$ 1.12  	&	 --  	&	 1.27 $\pm$ 0.11	&	 --  &	 1.68 $\pm$ 0.10 &	  -- 	&	 4.16$:$ 	\\

\hline															
Code	&	\multicolumn{2}{c}{\sc PyNeb}	&		&		&		&		&		\\
Aperture$^{(*)}$	&	\multicolumn{2}{c}{Integrated}	&		&		&		&		&		\\
ICF Scheme$^{(**)}$	&	\multicolumn{2}{c}{RR05}	&		&		&		&		&		\\
$N_e$	&	300~cm$^{-3}$	&	10$^5$~cm$^{-3}$	&		&		&		&		&		\\
\hline															
Fe$^{++}$/H$^+$	($\times10^{6}$) &	 0.58 $\pm$ 0.13	&	 0.56 $\pm$ 0.12	&		&		&		&		&		\\
ICF(Fe)	&   2.28 $\pm$ 0.31	&	2.28 $\pm$ 0.31	&		&		&		&		&		\\
Fe/H ($\times10^{6}$) &	1.31 $\pm$ 0.23	& 1.27 $\pm$ 0.23	&		&		&		&	 4.0	&	 --	\\

\hline

\multicolumn{8}{c}{$^{(*)}$ 2D integrated or slit; see text for details. $^{(**)}$ ICF schemes from \cite{kingsburgh94} (KB94),}\\

\multicolumn{8}{c}{\cite{Delgado2014} (DI14), \cite{Liu_etal_2000} (L00), and \cite{Rodriguez2005} (RR05); for He}\\

\multicolumn{8}{c}{no ICF is used. (:) Uncertain values.} \\ 


\end{tabular}
\end{table*}

\subsection{Physical and chemical properties} \label{sec:phys}

Table~\ref{tab:diagNP108} shows the gas diagnostic results obtained with the code {\sc NEAT} \citep[{Nebular Empirical Analysis Tool, version 2.2.40};][]{Wesson12} using the integrated spectrum line fluxes of Hen~2-108 as input. For the calculation of physical conditions and of chemical abundances, we considered only lines that were clearly from the nebula and not totally from the central source. The lines not considered were: N~{\sc ii} 4621, 4630; N~{\sc iii} 4634; C~{\sc iii} 4650, 4651 and 5696. These lines are well reproduced by the CS model discussed in Section \ref{sec:CSspec}. The code was executed using 5000 iterations in order to obtain uncertainties through a Monte-Carlo procedure. The table shows the values of the electronic density estimated from low and medium ionization ions, [\ion{S}{ii}] and [\ion{Cl}{iii}], respectively, as well as the  electronic temperature obtained from low ionization ion [\ion{N}{ii}] and middle ionization ion [\ion{O}{iii}]. For comparison, we also include in the table calculations from \citet{Gorny14} which use a methodology for the diagnostics similar to ours, but based on long-slit observations.

The extinction coefficient is obtained from H$\alpha$/H$\beta$. We used the extinction law of \citet{Cardelli89}. The extinction coefficient of this PN was determined by several authors and the results are consistent with our value, even though the methods and data are distinct. In addition to the extinction coefficient calculated by \citet{Gorny14} given in Table~\ref{tab:diagNP108}, \citet{Torres-Peimbert77} inferred $c$(H$\beta$)~=~0.4 and \citet{Pottasch11} $c$(H$\beta$)~=~0.53.

Table~\ref{tab:abundNPs} shows the ionic and elemental abundances obtained from the VLT/VIMOS data with the code {\sc NEAT}. The table provides abundances of He and N obtained from recombination lines, as well as abundances of N, O, S, Ar, Cl, and Fe. The total abundances were obtained using the ionization correction factors of \citet{DI14}. Abundances from the literature are also included for comparison. A full comparison taking into consideration the field of view used in each study is presented in Section~\ref{sec:module_specific_slits}.

The O/H and N/H abundances were derived from recombination and collisionally excited lines, which allowed us to estimate an abundance discrepancy factors ADF(O/H) = $9 \pm 3$ and ADF(N/H) = $22 \pm 15$ respectively. This result should be considered with care, as only a few lines of both ions are detected and used in obtaining the recombination abundance and for N there are no ions which emit both recombination and collisionally-excited lines in the optical. In particular the N/H ADF is likely to be overestimated due to residual contamination from emission lines from the central source (see section \ref{sec:CSspec} for details). Also, according to \cite{EMG12}, the nitrogen spectrum can be produced by continuum fluorescence, which may be specially important in low excitation conditions such as the ones for Hen~2-108. These abundances and respective ADF values should be considered as an order of magnitude estimates only. With the O and C recombination abundance determinations, albeit with low signal-to-noise ratio, it was possible to estimate C/O = $0.22 \pm 0.05$. 

\begin{figure}
\centering
\includegraphics[width=8.25cm]{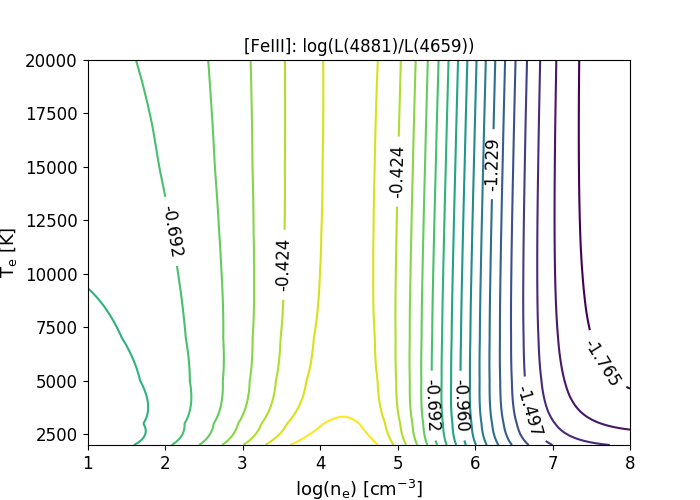}
\includegraphics[width=8.25cm]{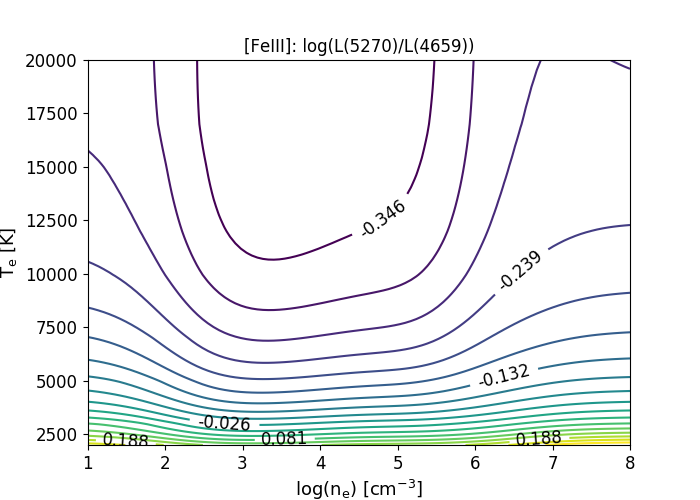}
\includegraphics[width=8.25cm]{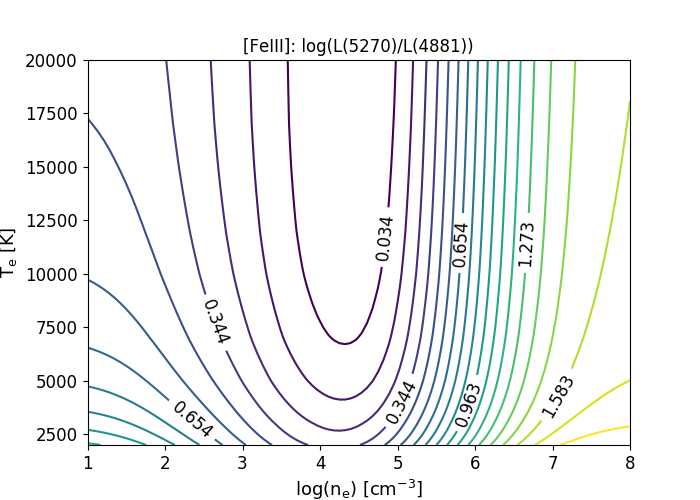}
\caption{[Fe~{\sc iii}] diagnostic diagrams for three line ratio indicators: $\lambda$4881/$\lambda$4659 (upper panel), $\lambda$5270/$\lambda$4659 (middle panel) and $\lambda$5270/$\lambda$4881 (lower panel), for $T_{\rm e}$ in the range from 2000 to 20000~K and log($N_{\rm e}$) in the range from 1 to 8.}
\label{FeIIIDD}
\end{figure}

We also determined the Fe abundance in Hen~2-108 from the [\ion{Fe}{iii}] $\lambda$4659, $\lambda$4881, $\lambda$5270 emission lines using the {\sc pyneb} package \citep{Luridiana15} since {\sc NEAT} did not perform this estimate. Since [\ion{Fe}{iii}] line ratios can also be used as density indicators \citep{Laha2017}, the ($T_{\rm e}$-$N_{\rm e}$) diagnostic diagrams are presented in Fig.~\ref{FeIIIDD} for each of three lines. The observed line ratios are log[\ion{Fe}{iii}]~$\lambda$4881/$\lambda$4659=-0.72$\pm$0.07, log[\ion{Fe}{iii}]~$\lambda$5270/$\lambda$4659=-0.36$\pm$0.06, and log[\ion{Fe}{iii}]~$\lambda$5270/$\lambda$4881=0.37$\pm$0.08. Comparing the observed ratios to the values in the top and bottom panels, two densities are possible, $N_{\rm e}\sim$100~cm$^{-3}$ or $\sim$10$^{5.5}$~cm$^{-3}$, respectively. From the middle panel, the observed [\ion{Fe}{iii}]~$\lambda$5270/$\lambda$4659 ratio is valid for $T_{\rm e}>$12\,000~K but considering the uncertainty of the line ratio, Te can be as low as $\sim$8000K and $N_{\rm e}$ ranges from 10$^{2.5}$ to 10$^{5.5}$~cm$^{-3}$.
Overall, these results indicate the presence of a gas with $N_{\rm e}$ between 10$^3$ and 10$^5$~cm$^{-3}$ in the central region of the nebula, where most of the iron emission is produced (Fig.~\ref{fig:flux108-2}).

The ionic abundance of the doubly ionized Fe is computed 5.8$\times$10$^{-7}$ for $N_{\rm e}$=300~cm$^{-3}$ and 5.6$\times$10$^{-7}$ for $N_{\rm e}$ =10$^5$~cm$^{-3}$. The elemental abundance of Fe is between 1.31$\times$10$^{-6}$ and 1.27$\times$10$^{-6}$ using the ICF expressions provided by \cite{Rodriguez2005}. The uncertainties were computed considering a Monte Carlo approach with 200 trials. Note that our ionic abundance of Fe is found to be close to the value computed by \cite{Pottasch11} but our elemental Fe abundance is three times higher because of the difference in ICF(Fe) between the two studies.

\section{Satellite code}
\label{sec:satellite}

The spatially resolved study of the chemistry in PNe is complicated by the fact that the usual empirical procedures used to obtain abundances do not take into account the spatial complexity of the nebulae. The Spectroscopic Analysis Tool for intEgraL fieLd unIt daTacubEs ({\sc satellite}\footnote{{\sc satellite} v.1.3 is publicly available on Github: \url{https://github.com/StavrosAkras/SATELLITE}}; \citealt{Akras22}; \citealt{Akras20}) was developed to perform a series of calculations to assist in the analysis of the spatial variation on typical nebular diagnostics and abundance determination. The code has four different modules: \textit{rotational analysis}, \textit{radial analysis}, \textit{specific slit analysis}, and \textit{2D spatial analysis} of the emission line flux maps generated from the data cubes of any IFU. The code simulates pseudo-slits spectra for a direct comparison with the data from long-slit spectroscopy as well as the construction of emission line ratios diagnostic diagrams (i.e. SMB, \citealt{Sabbadin1977}; BPT, \citealt{Baldwin1981}; VO, \citealt{Veilleux1987}). All the physical parameters (c(H$\beta$), $T_\textrm{e}$ and $N_\textrm{e}$, ionic and total abundances, and ionization correction factors (ICFS)) are computed for each pseudo-slit through the implementation of the {\sc pyneb} package 1.1.15 \citep{Luridiana15}. Atomic data available in {\sc pyneb} can be selected by the user of {\sc satellite}. For the case of Hen2-108, the atomic data from Chianti database \citep{Dere1997,DelZanna2021} were used for direct comparison with the results obtained with {\sc neat}. In the following subsections, we discuss the application of the different modules and capabilities of {\sc satellite} to the Hen~2-108 VIMOS data.

\subsection{Rotation analysis module}

\begin{figure*}
\centering
\includegraphics[width=8.5cm]{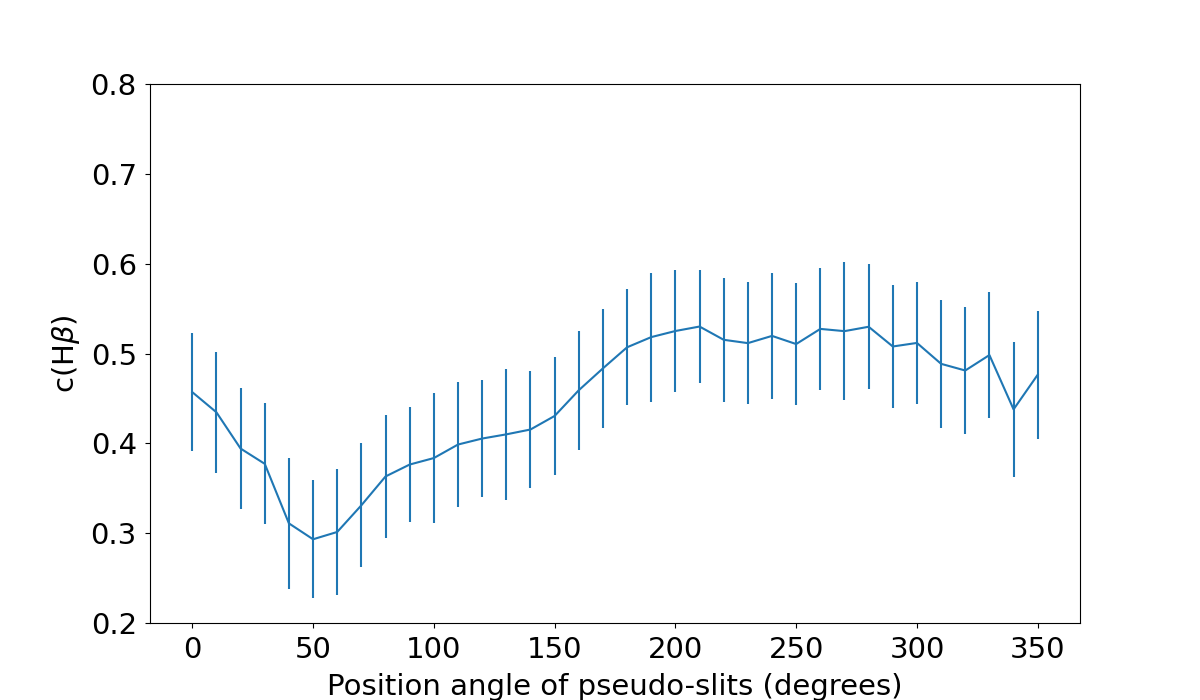}
\includegraphics[width=8.5cm]{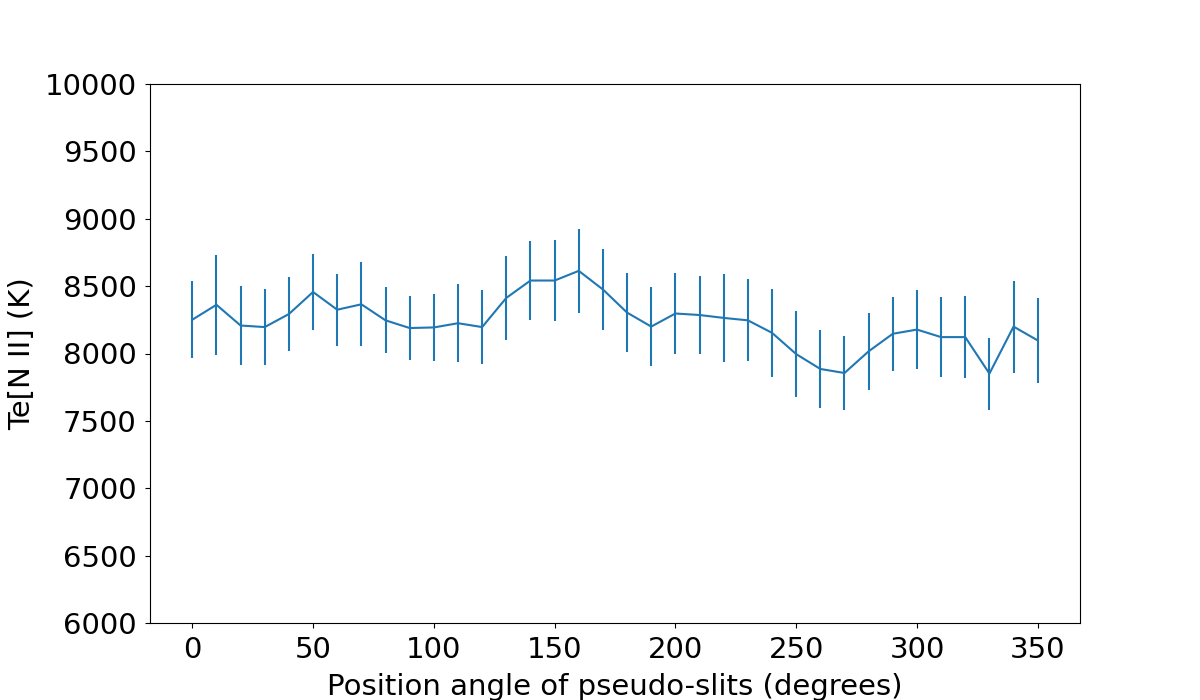}
\includegraphics[width=8.5cm]{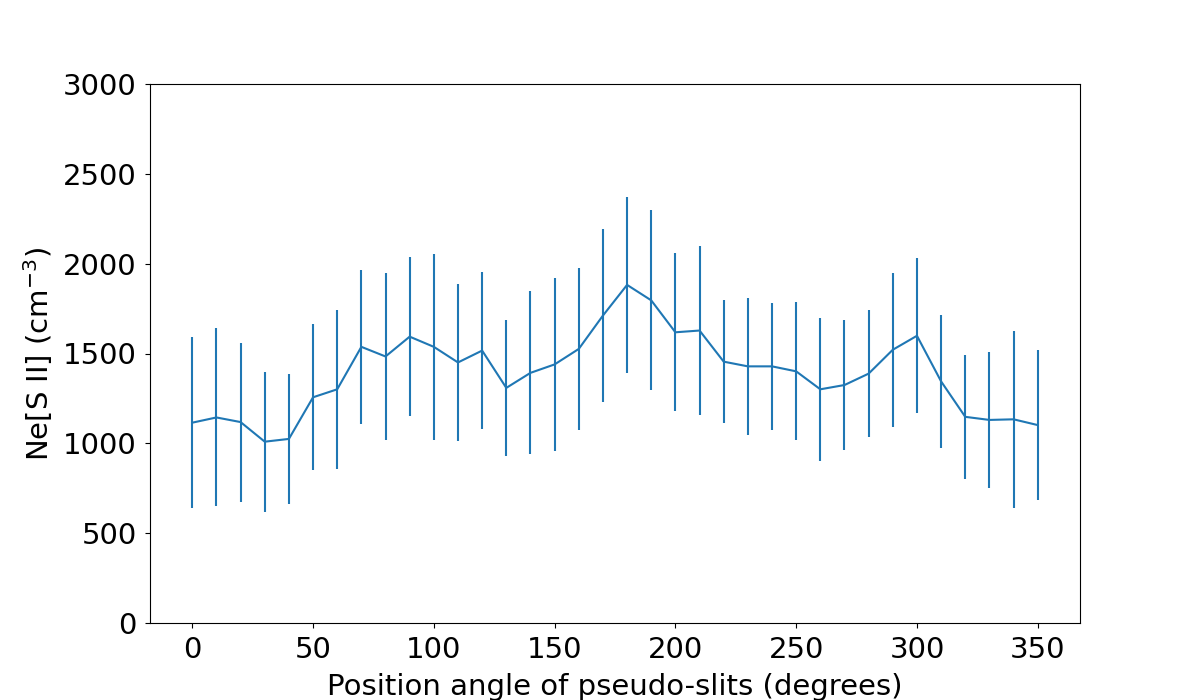}
\includegraphics[width=8.5cm]{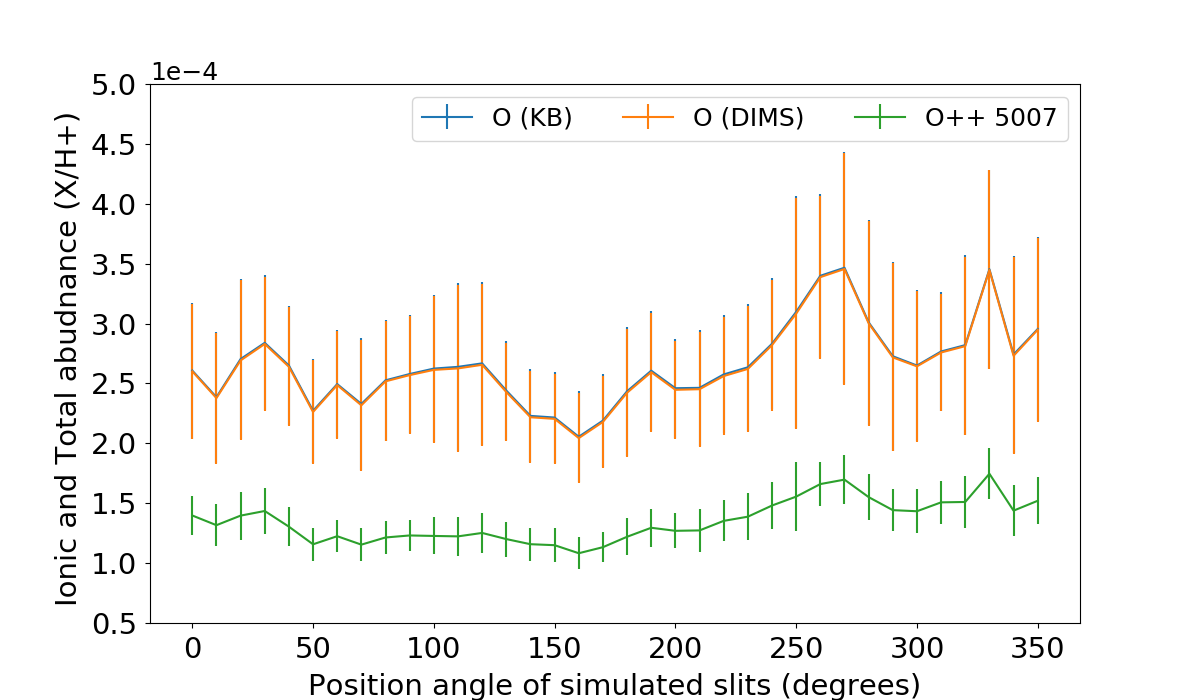}
\includegraphics[width=8.5cm]{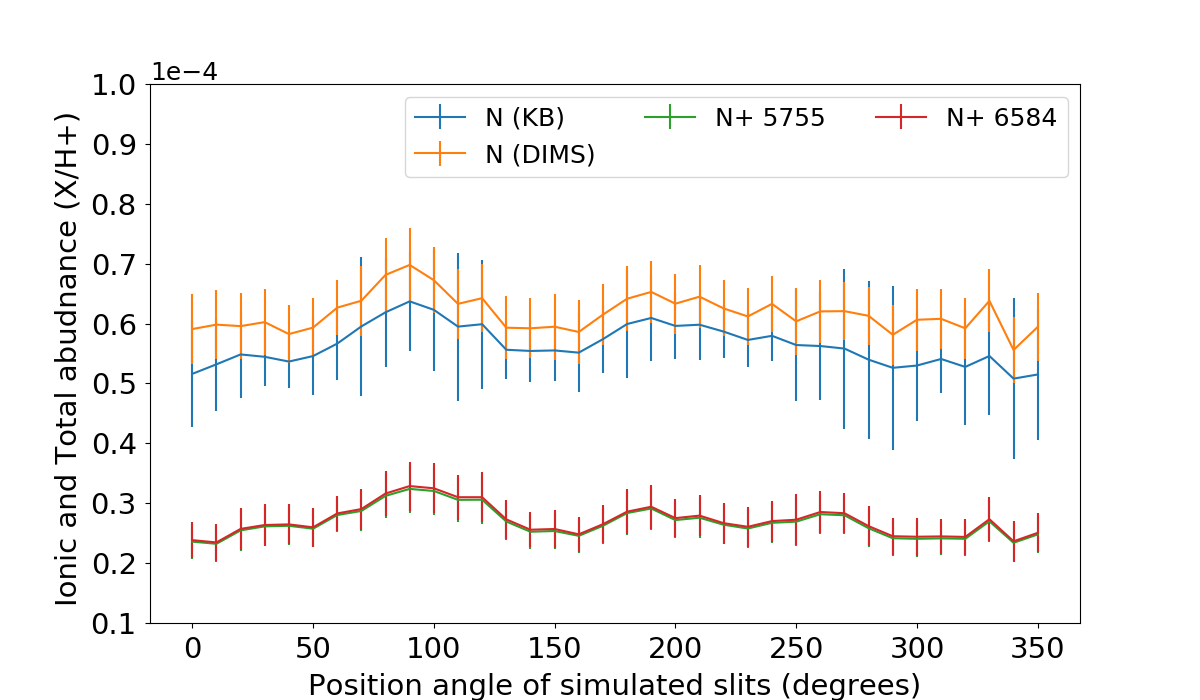}
\includegraphics[width=8.5cm]{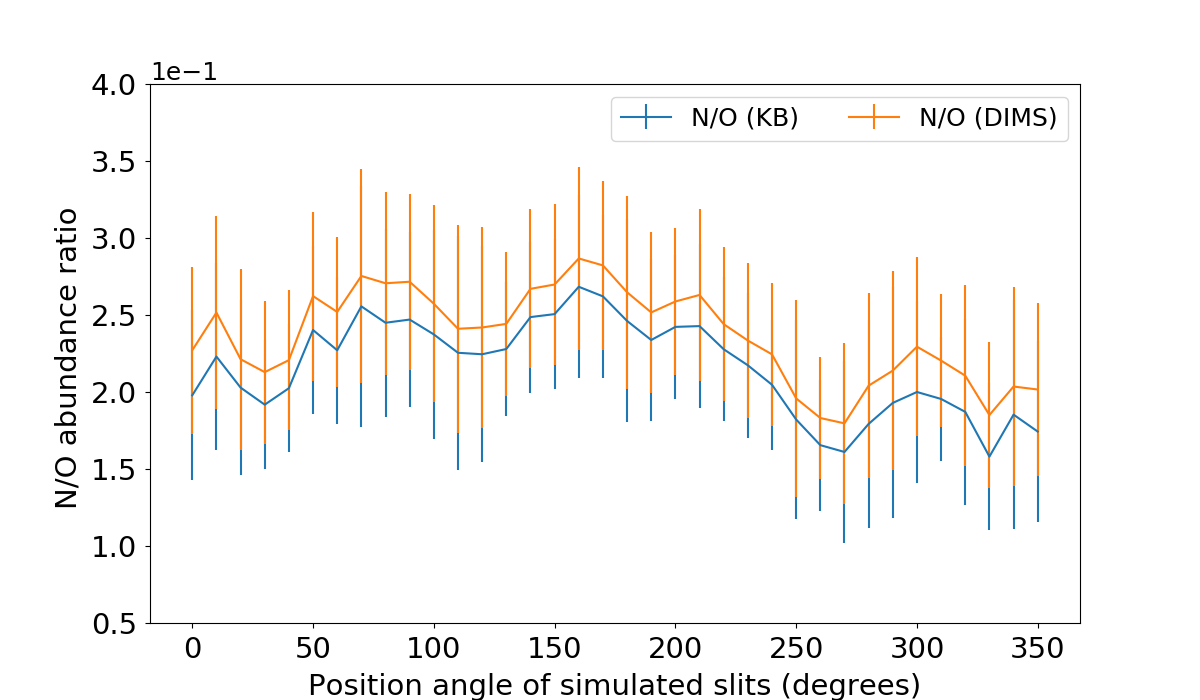}
\caption{Rotation analysis: Left panels: $c$(H$\beta$) (top), $T_\textrm{e}$ (middle) and $N_\textrm{e}$ (bottom), right panels: Ionic and total abundances of O (top), N (middle) and N/O (bottom) ratio as function of the pseudo-slits' PA. A pseudo-slit oriented north–south has PA=0 degrees and increases rotating to the east.}\label{anglechbTeNe}

\label{figabund}
\end{figure*}

The \textit{rotation analysis module} simulates a number of slits placed radially from the centre of the observed nebula, varying the position angle (PA) from 0 to 360~degrees. In this work, we considered pseudo-slits with width and length of 5 and 20 spaxels or equivalently 3.35 and 13.4~arcsec, respectively. Figure~\ref{anglechbTeNe} displays the variation of the extinction coefficient, $T_\textrm{e}$ and $N_\textrm{e}$ (obtained employing the [N~{\sc ii}] and [S~{\sc ii}] diagnostics, respectively) as functions of the pseudo-slits' PA. A non-negligible variation in $c$(H$\beta$) is found. $c$(H$\beta$) starts with a value of 0.45 at PA=0 (north-south direction), decreases down to 0.3 at PA=50~degrees and increases again to 0.5 for PAs between 170 and 330~degrees. On the other hand, $T_\textrm{e}$ and $N_\textrm{e}$ do not show significant variation with the PA and their median values are 8217~K and 1415~cm$^{-3}$ with standard deviations of 180~K and 215~cm$^{-3}$, respectively.

Ionic and total chemical abundances of O and N as well as the N/O ratio as a function of the pseudo-slits' PA are displayed in Fig.~\ref{figabund}. No significant variation in the elemental abundances with the direction of the pseudo-slits is found. Note that for N and S (not shown here), there is a small difference between the elemental abundances obtained using the ICFs formulae from \cite{kingsburgh94} (blue line) and \cite{Delgado2014} (orange line).

The results of this analysis indicate that the elemental abundances' distribution shows some deviations from spherical symmetry, albeit with large uncertainties. The error bars of $c$(H$\beta$), $T_\textrm{e}$ and $N_\textrm{e}$ in Fig.~\ref{anglechbTeNe}, as well as the errors of all the physical parameters determined by {\sc satellite}, correspond to the standard deviations from a Monte Carlo distribution of the emission line intensities assuming 200 replicates.

\subsection{Radial analysis module}

\begin{figure*}
\centering
\includegraphics[width=8.25cm]{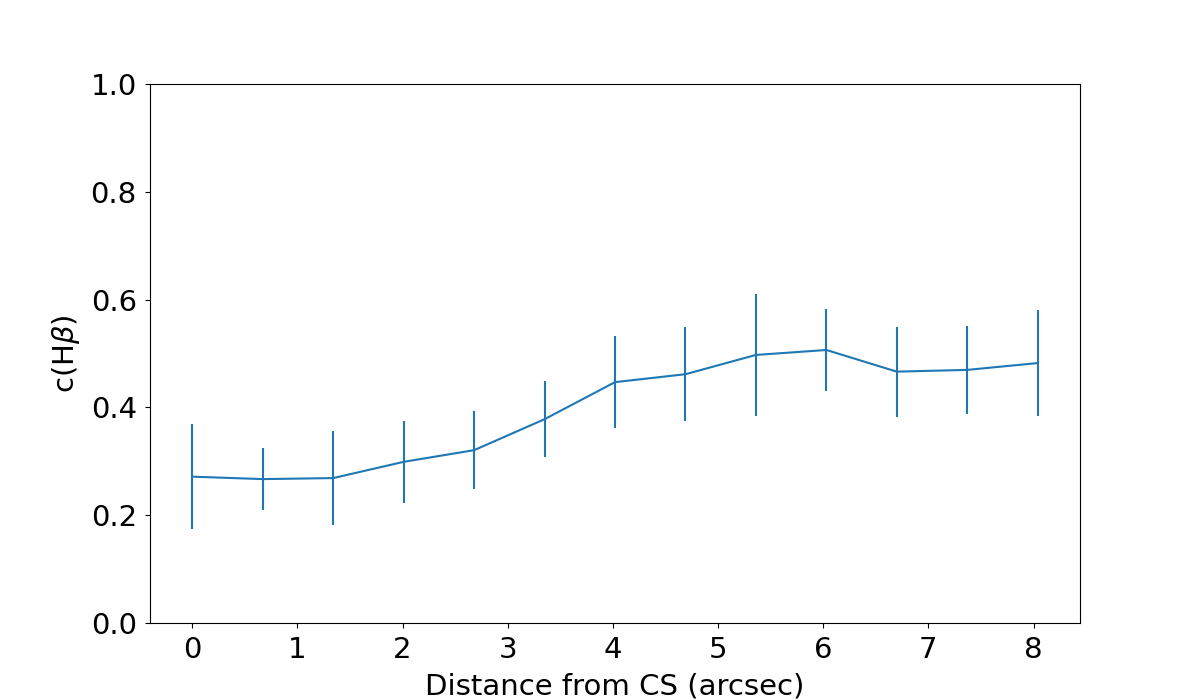}
\includegraphics[width=8.25cm]{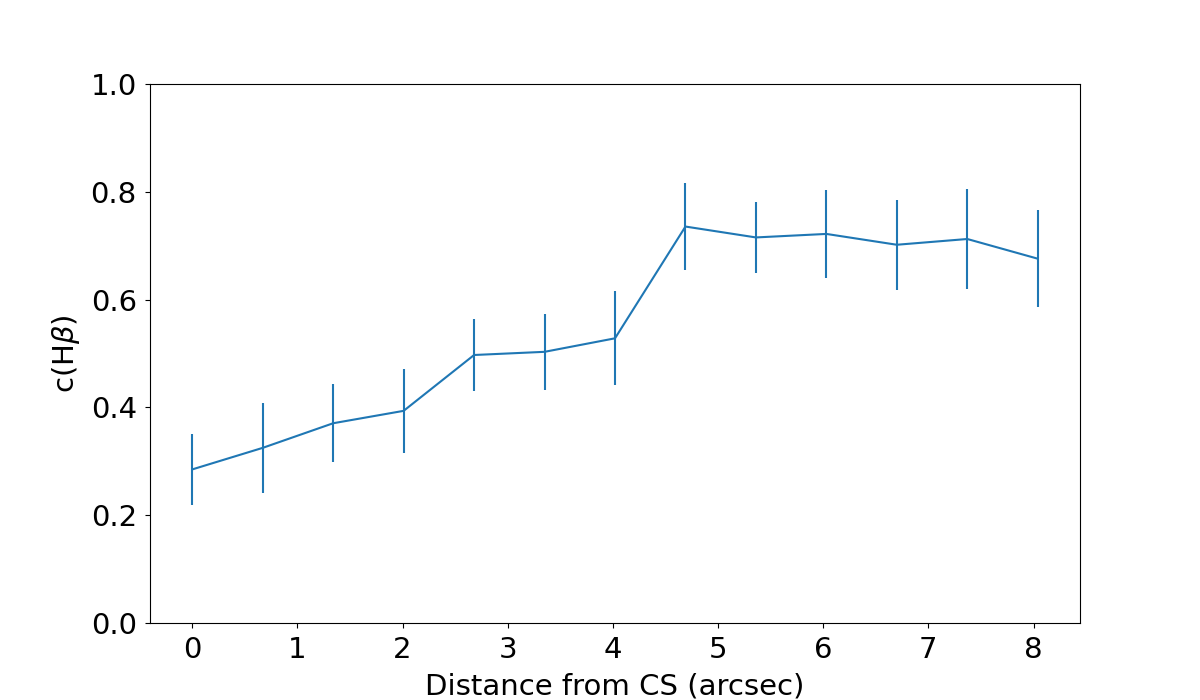}
\includegraphics[width=8.25cm]{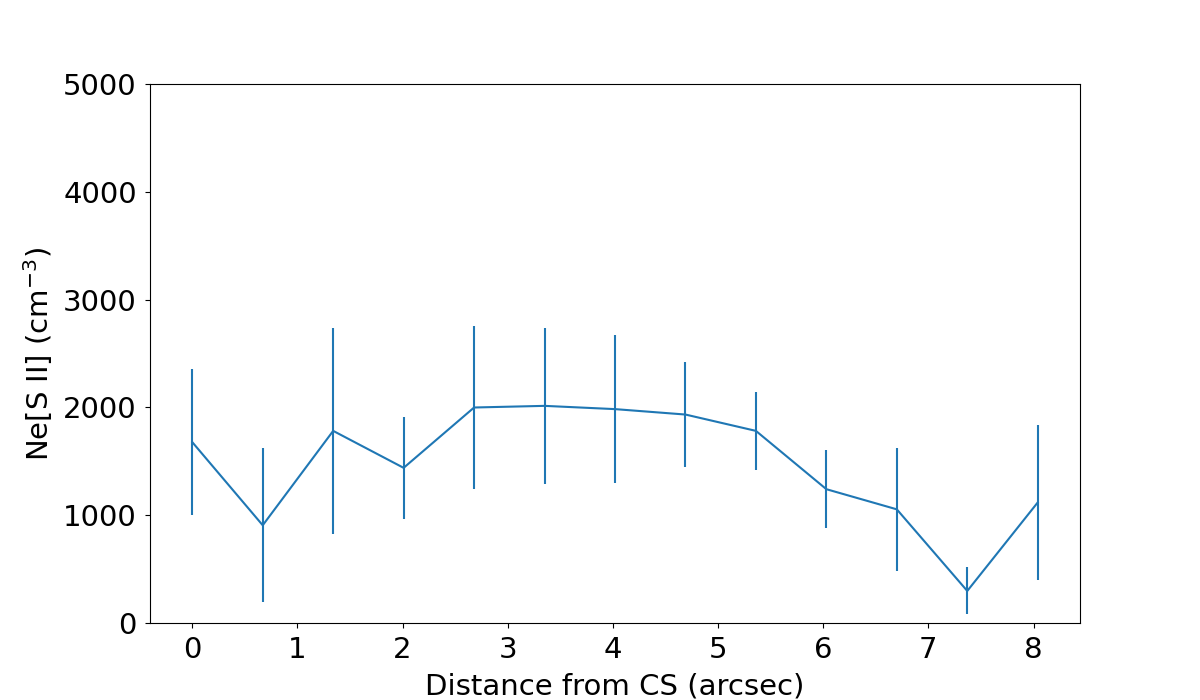}
\includegraphics[width=8.25cm]{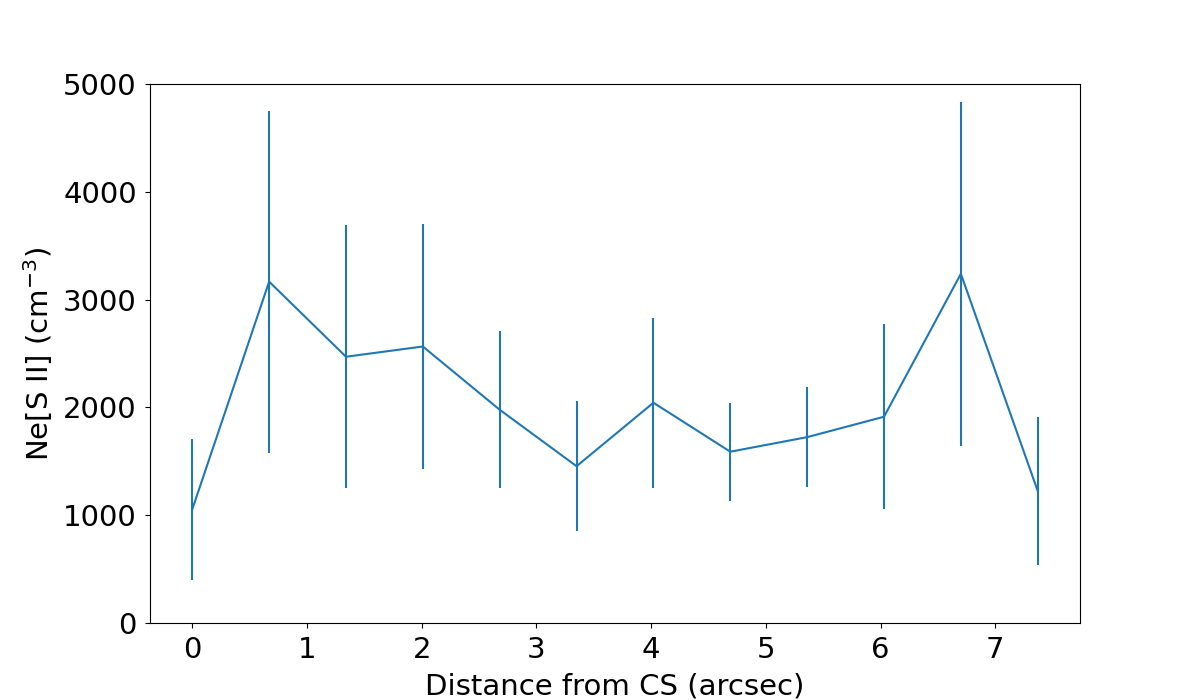}
\includegraphics[width=8.25cm]{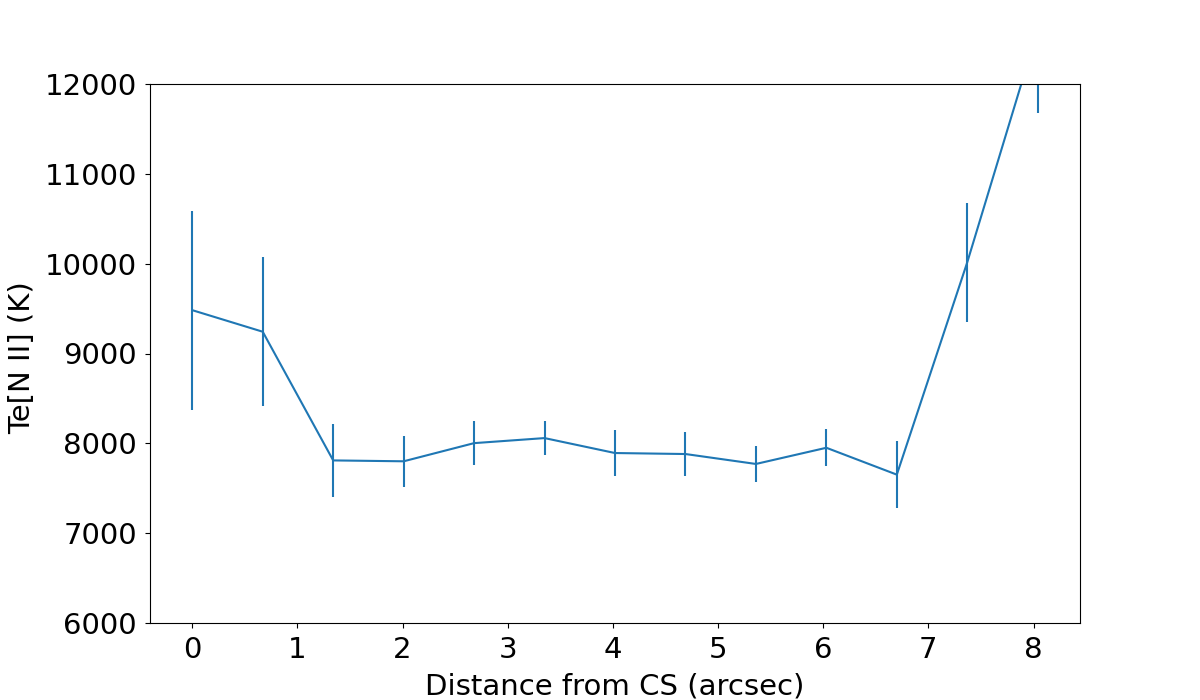}
\includegraphics[width=8.25cm]{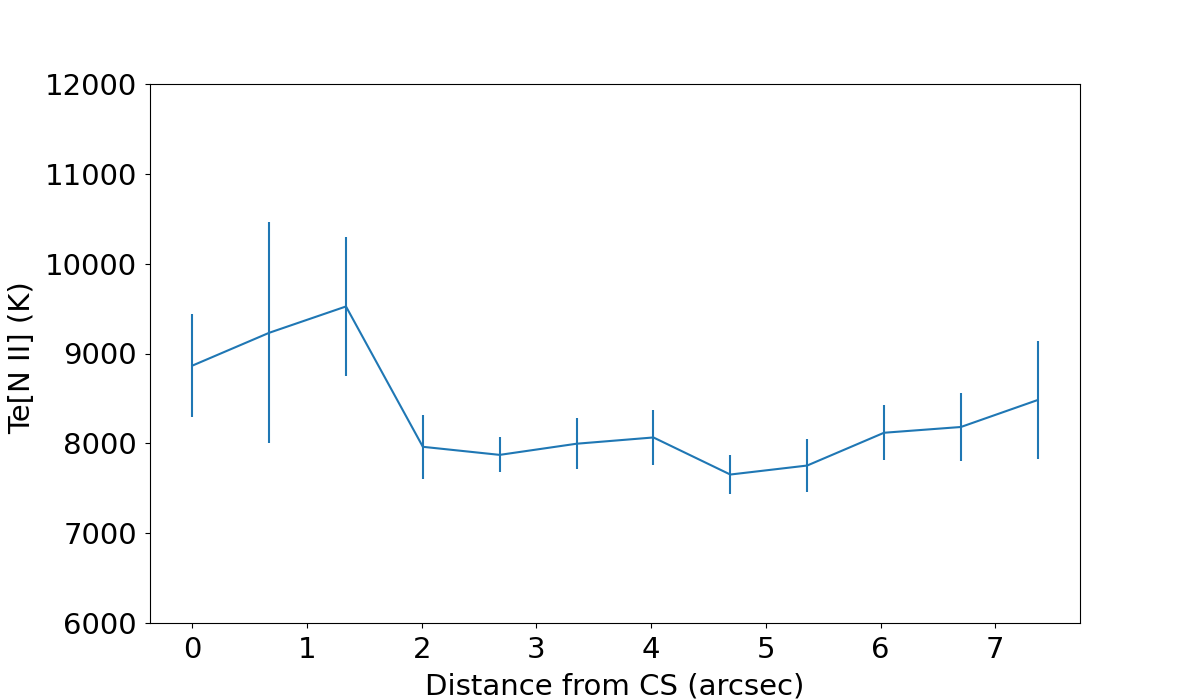}
\caption{Radial analysis: $c$(H$\beta$) (upper panel), $N_\textrm{e}$ (middle panel) and $T_\textrm{e}$ (lower panel) as a function of the distance from the central star for a pseudo-slit at PA=90~degrees (left panels) and PA=180~degrees (right panels).}
\label{radialchbTeNe}
\end{figure*}

The {\it radial analysis module} performs a spectroscopic analysis in a radial direction for a specific pseudo-slit. The direction (PA), width and length of the pseudo-slit are chosen by the user and the products of the analysis are provided as functions of the distance from the CS. For the case of Hen~2-108, we applied this module for two directions: PA=90~degrees (eastern direction) and PA=180~degrees (southern direction). The same slit width and length as in {\it the rotation analysis module} were adopted. The variation of $c$(H$\beta$), $T_\textrm{e}$, and $N_\textrm{e}$ as a function of the distance from the CS is presented in Fig.~\ref{radialchbTeNe}. The step in these plots is one spaxel or 0.67~arcsec. 

For both pseudo-slits, $c$(H$\beta$) is found to be as low as 0.3 in the inner part of the nebula and increases outwards, reaching a value of 0.5 for the eastern part and $\sim$0.7 for the southern part. Note that the radial analysis with the {\sc satellite} code results in very low $c$(H$\beta$) for the inner region of the nebula, while the map in Fig~\ref{fig:extinction} displays values as high as 0.7-0.8. This is caused by the difference in the method. For the map, both the \ha/H$\beta$~and~H$\gamma$/H$\beta$ lines ratios are used to get the extinction coefficient, while the {\sc satellite} considers only the \ha/\hb~ratio.

$T_\textrm{e}$ and $N_\textrm{e}$ display variations with the distance from the CS although with large uncertainties, especially in the inner regions where the signal-to-noise ratio is low for the used lines. Note that, for $r<$1.6~arcsec, $T_\textrm{e}$ is higher in the inner region of the nebula, which is consistent with the 2D maps (Fig~\ref{fig:denstemp-He2-108}).

\subsection{Specific slits analysis module} \label{sec:module_specific_slits}

The third module of {\sc satellite}, the {\it specific slits analysis module}, deals with the simulation of ten pseudo-slits at specific positions and orientations on the nebula for a direct comparison with long-slit spectroscopy studies.

The long slit observation from \citet{Gorny14} is simulated considering a 3 spaxels (= 2~arcsec) wide pseudo-slit. A second pseudo-slit 25 spaxel wide covering the whole nebula is used to obtain the integrated spectrum of Hen~2-108. Line intensities from both spectra are presented in Table~\ref{tab:fluxlinesNP}. The integrated fluxes obtained with {\sc satellite} (including the absolute H$\beta$ flux of 1.9$\times$10$^{-12}$) and {\sc neat} are also in very good agreement. The discrepancy between the {\sc satellite} analysis and the results from \citet{Gorny14} is less than 5~percent for the majority of the emission lines, except the He~{\sc ii} and [S~{\sc iii}] lines. Note that the difference between the integrated and pseudo-slit He~{\sc ii} intensities is significant. The long-slit used by \citet{Gorny14} covers only a small part of the nebula towards the centre, where the He~{\sc ii} emission emanates. Part of the H$\beta$ flux is, however, not included, resulting in a higher He~{\sc ii}/H$\beta$~ratio compared to the integrated spectrum value.

All physical parameters ($c$(H$\beta$), $T_\textrm{e}$, $N_\textrm{e}$, abundances and ICFs) computed by {\sc satellite} for the integrated and G14's pseudo-slit spectra are given in Tables~\ref{tab:diagNP108} and \ref{tab:abundNPs}.

\subsection{2D analysis module}

\begin{figure}
\centering
\includegraphics[width=8.0cm]{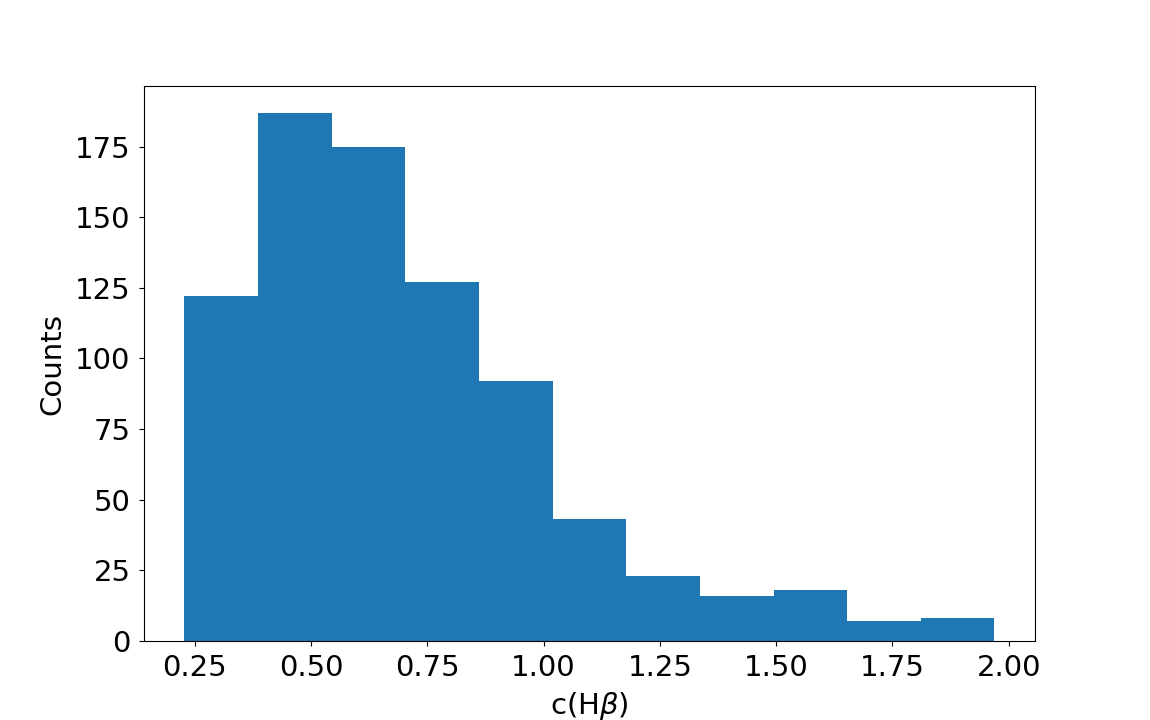}
\includegraphics[width=8.0cm]{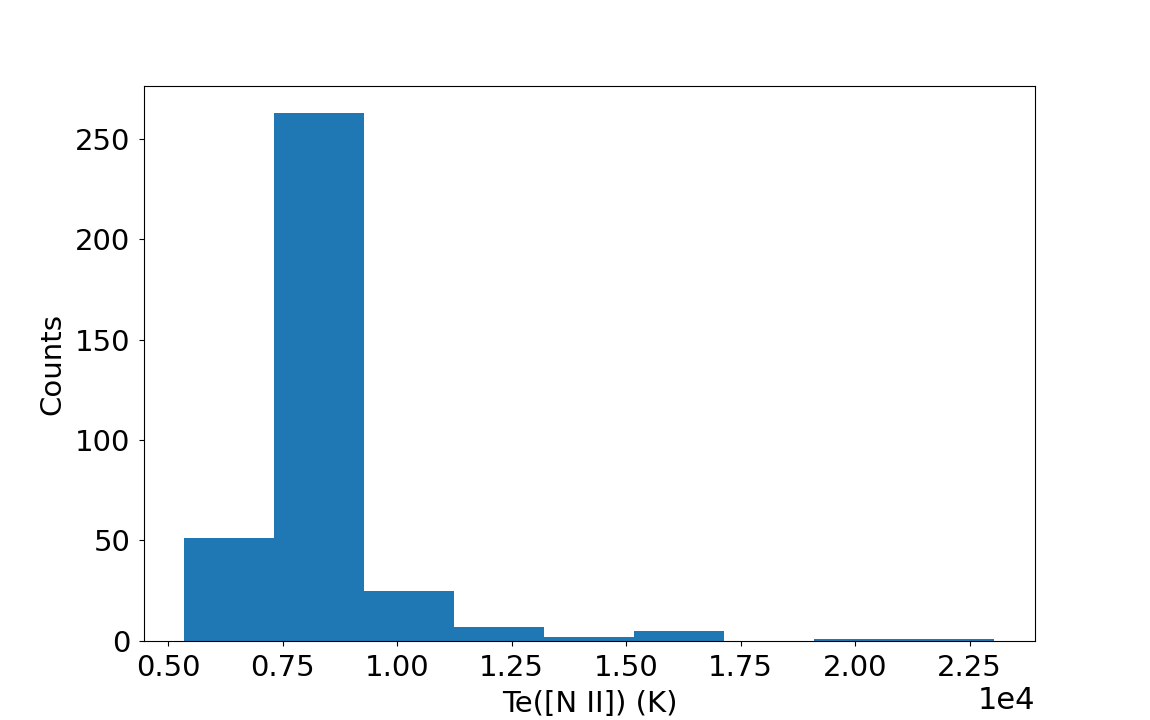}
\includegraphics[width=8.0cm]{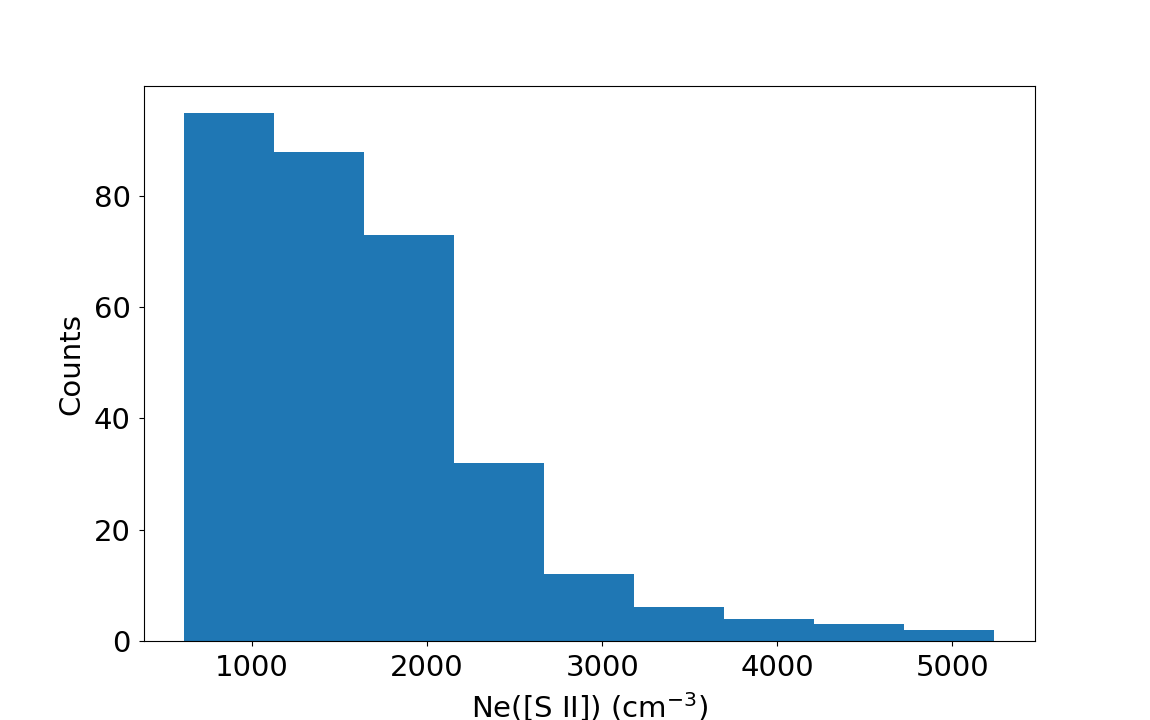}
\caption{The distribution of the $c$(H$\beta$), $T_\textrm{e}$ (in 10$^4$~K), and $N_\textrm{e}$ (in cm$^{-3})$ in their corresponding 2D maps.}
\label{histchbTeNe}
\end{figure}

{\sc satellite} also calculates all the physical parameters for each individual spaxel and provides maps for all of them, as well as a number of line ratio maps defined by the user. The maps produced by {\sc satellite} are similar to what is shown in Figs.~\ref{fig:flux108-2} and are therefore not shown here. 

Histograms of the nebular parameters are also provided by {\sc satellite} for a better illustration of their distributions. The histograms of $c$(H$\beta$), $T_\textrm{e}$, $N_\textrm{e}$ for Hen~2-108 are shown in Fig.~\ref{histchbTeNe}. Most of the spaxels have $c$(H$\beta$) between 0.2 and 0.8 with a peak at $\sim$0.5. This range is consistent with the values discussed in previous sections. Note that the histogram indicates values higher than 1.0 which are not observed in the map of Fig.~\ref{fig:extinction} because of the mask of S/N=6 that applied on the datacube for the construction of the line maps. The distributions of $T_\textrm{e}$ and $N_\textrm{e}$ are concentrated in the ranges of 7500 to 8500~K and 500 to 2000~cm$^{-3}$, respectively.

\section{Spectrum of the central region}
\label{sec:CSspec}

The spectrum of the central region presents several differences with respect to the rest of the nebula. In this region, the CS can contribute significantly to the observed integrated spectrum. To assess the star's contribution and study which type best explains the spectral characteristics, we extracted the spectrum of the central region of the 2:1 magnification data cube and subtracted the nebular contribution. This was achieved by fitting two-dimensional Gaussians plus a second degree surface polynomial to the central region of the PN, defined as a 5~arcsec width centred at the star, at each point along the dispersion axis.

\begin{figure*}
    \centering
    \includegraphics[width=\textwidth]{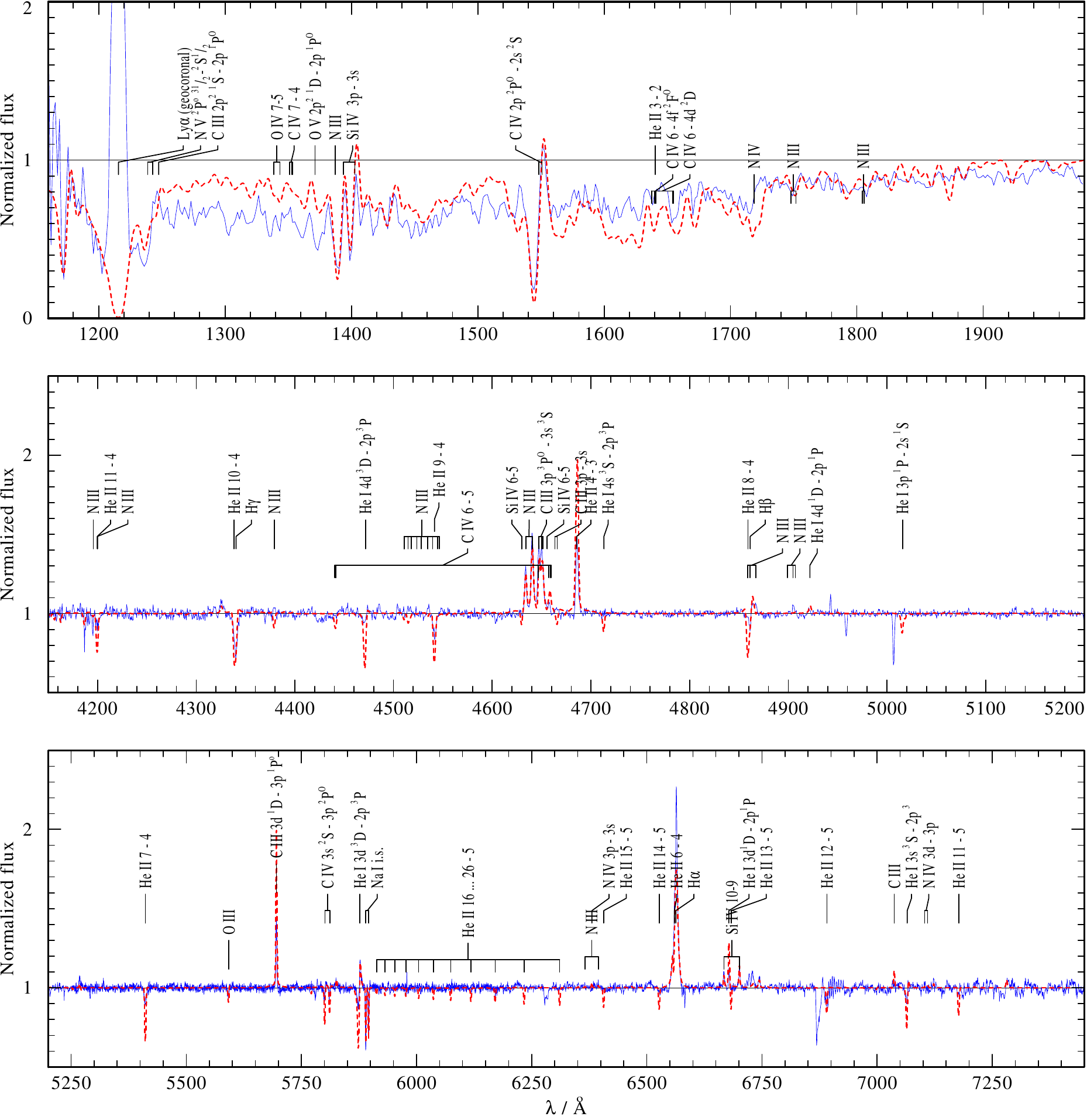}
    \caption{Details of the UV and optical spectrum of the central star Hen\,2-108: observations (blue solid lines) plotted vs.\ the best fitting \textit{PoWR} model (red dashed). In the optical range, the model
 spectrum was convolved with a Gaussian with $1.6\,$\AA\ FWHM to match
 the resolution of the observation, inferred from the interstellar
 Na\,{\sc i} doublet. For the UV range a Gaussian with FWHM of 5\,\AA{} is used, 
 according to the low dispersion \textit{IUE} observations.}
    \label{fig:CSspec}
\end{figure*}

\begin{figure*}
\centering
\includegraphics[width=\textwidth]{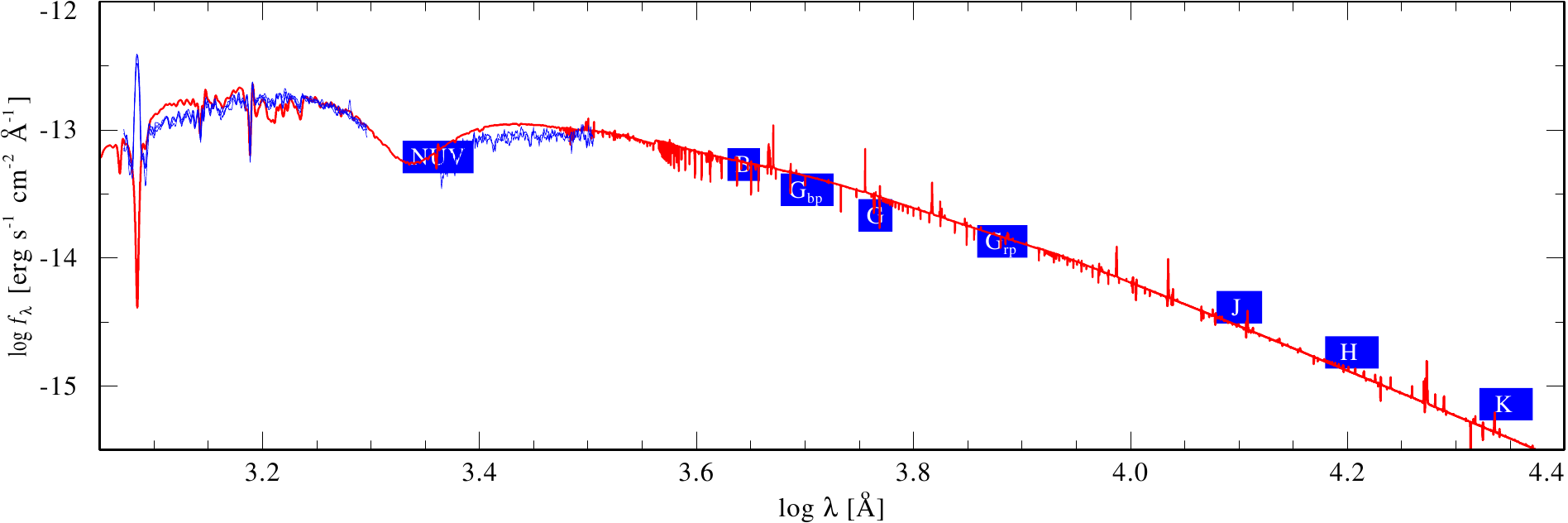}
\caption{Spectral energy distribution of the central star: \textit{PoWR} model (red line) shown together with \textit{Galex} NUV, \textit{Gaia}, Johnson B and JHK$_\text{S}$ photometry (blue boxes)
and \textit{IUE} spectra (blue lines).}
\label{fig:CSsed}
\end{figure*}

\begin{table}
    \centering
    \caption{Parameters of the CS from our analysis with \textit{PoWR}}
    \begin{tabular}{lcl}
 \hline
Parameter & Value & Comment\\
\hline
$d$ [kpc]                            & $3.64^{+0.30}_{-0.23}$       & \cite{Bailer-Jones2021} \\
$E(B-V)$ [mag]                       & $0.37\pm0.03$     & SED fit \\
$M_{\star}$ [M$_\odot$]              & 0.5               & adopted \\
$T_\star$ [kK]                       & $41.5\pm2.0$   & Defined at $\tau_\text{Ross}=20$ \\
$\log(L_{\star}$/L$_\odot)$            & $3.62^{+0.10}_{-0.14}$       &  \\
$R_{\star}$ [R$_\odot$]              & $1.25\pm0.22$     & stellar radius \\ 
$\dot{M} [$M$_{\odot}$~yr$^{-1}]$      & $6.0^{+2.2}_{-1.6}\times10^{-8}$            &  \\ 
$\log R_\text{t}$ [R$_\odot$]        & $1.83\pm0.02$     & transformed radius\\
$\varv_{\infty}$ [km\,s$^{-1}$]      & $1200\pm200$      &  \\
$D$                                  & 4                & density contrast \\
$\beta$                              & 4                 & $\beta$-law exponent\\
\hline
\multicolumn{3}{l}{Chemical abundances (mass fraction)}\\
\hline
H  &  $0.50^{+0.24}_{-0.10}$          & \\[0.05cm]
He &  $0.48^{+0.10}_{-0.24}$    & \\[0.05cm]
C  &  $7.1^{+4.7}_{-2.4}\times10^{-3}$     & 3$\times$ solar \\[0.05cm]
N  &  $1.4^{+0.7}_{-0.4}\times10^{-3}$     & 2$\times$ solar\\[0.05cm]
O  &  $5.73^{+0.3}_{-2.8}\times10^{-3}$    & solar\\[0.05cm]
Si &  $6.65^{+0.7}_{-0.7}\times10^{-4}$    & solar\\[0.05cm]
Fe &  $1.4\times10^{-3}$     & solar \\
\hline
    \end{tabular}
    \label{tab:csparam}
\end{table}

As can be seen in Fig.~\ref{fig:CSspec}, the extracted spectrum shows clear absorption features such as \ion{He}{ii} $\lambda$4200, H$\gamma$, \ion{He}{i} $\lambda$4471, \ion{He}{ii} $\lambda$4541, \ion{He}{ii} $\lambda$5412, \ion{He}{i} $\lambda$5876, which shows a clear P~Cygni type profile and Na~I doublet $\lambda\lambda$5890,5895. We also detected absorption features at 5780, 5797 and 5812~\AA. Two of these features, 5797 and 5812~\AA, were identified as the \ion{C}{iv} doublet by \citet{Marcolino03}. Those authors, however, do not mention the very obvious feature at 5780~\AA\ in their spectrum. The three features coincide with the positions of lines known to be of diffuse interstellar bands (DIBs). These features were discovered by \citet{Heger22} and confirmed by \citet{Merrill36}. Further details about their properties are discussed by \citet{Sarre06} and references therein. However, our spectrum is not of sufficient quality to rule out the 5797 and 5812~\AA~lines as the \ion{C}{iv} doublet. It is possible that there is a combination of both DIBs and \ion{C}{iv} doublet in these absorption lines. A higher resolution and signal-to-noise spectrum is necessary to confirm this.

Also confirmed here by the extracted central region spectrum, the lines of \ion{N}{iii} $\lambda$4634, \ion{He}{ii} $\lambda$4686 and \ion{C}{iii} $\lambda$5696 are mostly stellar in origin, although from our data we can not specify exactly the proportion. The lines of \ion{C}{iii} $\lambda$5696, \ion{C}{ii} $\lambda$7236 and \ion{He}{ii} $\lambda$4686 were identified by \citet{Gorny14} as of stellar origin.

\citet{Marcolino03} emphasize that the lines of \ion{N}{iii} $\lambda4634$ and $\lambda4641$ are present in [WELS] and the first is less intense than the second. The first is seen in our spectrum next to the line \ion{N}{ii} $\lambda4630$, but $\lambda4634$ cannot be separated from $\lambda4639$, identified as \ion{O}{ii}. The same authors also identify the line $\lambda4658$ as \ion{C}{iv}, while we identified the same line as [\ion{Fe}{iii}]. Our detection is supported by the detection of other [\ion{Fe}{iii}] lines and no detection of other \ion{C}{iv} lines. It is seen in the flux map (Section~\ref{sec:spatialanalisys}) that the $\lambda4658$ wavelength is extended just beyond the central region, which may indicate that this emission is nebular. This region of the spectra can be very complex and can include many lines, as it can be seen for example in \citet{Peimbert04}. \citet{Weidmann15} disagree with the use of [WELS] as a classification type since many of the CSPNs were classified using low resolution spectra. 

Given the observed features such as the P~Cygni profiles and broad $H\alpha$ emission, together with the other emission lines seen in the CS of Hen~2-108, we calculated a set of models with the Potsdam Wolf-Rayet (\textit{PoWR}) code for expanding atmosphere to infer the stellar parameters, which are summarized in Table~\ref{tab:csparam}. 

Additionally to the optical spectrum of the CS, we use for our spectral analysis the coadded, low-dispersion \textit{IUE} observations sp47512 and sp47513, and also lp25380, all taken on 21. April 1993. 

The basic assumptions of the \textit{PoWR} code are spherical symmetry and stationarity of the 
flow. The radiative transfer equation is solved in the comoving frame of the expanding atmosphere, iteratively with the equations of statistical equilibrium  and radiative equilibrium. In the subsonic part,  the velocity field is implied by the hydrostatic density stratification according to the continuity equation. For the supersonic part of the wind, we prescribe the velocity field $\varv(r)$ by a so-called $\beta$-law, where the free parameter $\beta$ for WR stars is usually set to $\beta=1$ or $\beta=0.8$ for O-type stars.

For Hen\,2-108 we find that the widths of the optical lines correspond to expansion velocities of a few hundred km/s, while the P Cygni profile of the C\,{\scshape iv} resonance doublet in the \textit{IUE} spectrum indicates a terminal velocity of about $(1200\pm200)$\,km~s$^{-1}$. A consistent spectral fit of the widths of the C\,{\scshape iv} resonance doublet and the optical emission lines simultaneously is achieved for $\beta= 4$, i.e.\ a much shallower velocity law. We note that such a $\beta$ value is also used for the spectral fit of the qWR star HD\,45166 \citep[cf.][]{Groh2008}, that has a qualitatively similar spectrum.

Line broadening by microturbulence is also included in our models. From the shape of the line profiles, we deduce a microturbulence velocity of about $50\,\text{km}\,\text{s}^{-1}$.

We adopted a stellar mass of $0.5\,M_\odot$, which is slightly below the mean value for CSPNe of $0.6\,M_\odot$ \citep[see e.g.][]{schoenberner2005,miller-bertolami2007} and accounts for our findings of a progenitor with a mass below $1\,M_\odot$ (see Sect.~\ref{sec:conclusions}). Usually, the value of $M_\star$ has no noticeable influence on the synthetic wind spectra.

The stellar temperature $T_\star$ (defined at the continuum $\tau_\text{Ross}=20$) can be estimated from the relative strengths of spectral lines of different ionization stages of the same element. We used He\,{\scshape i} vs.\ He\,{\scshape ii}, C\,{\scshape iii} vs.\ C\,{\scshape iv}, and N\,{\scshape iii} vs.\ N\,{\scshape iv}. While the nitrogen lines are relatively well reproduced by our best fitting model and the fit to the carbon lines is acceptable, we could not find a model that results in a comparable fit quality to all of the helium lines. Models with lower temperatures of about $38\,$kK can reproduce the strengths of the He\,{\scshape ii}\,$\lambda$4686 emission line quite exactly, but result in too strong He\,{\scshape i} emission and N\,{\scshape iii} wind absorption lines. Moreover, such models result in a worse fit to the iron forest in the UV range. 

From the absolute strengths of the wind lines in the continuum-normalized spectrum we infer the transformed radius $R_\text{t}$ \citep[see][]{Schmutz1989}, a quantity which takes the invariance of the volume emission measure normalized to the stellar surface into account and is defined as

\begin{equation}
\label{eq:rt}
R_{\mathrm{t}} = 
R_\star \left(\frac{\varv_\infty}{2500 \, \mathrm{km}\,\mathrm{s^{-1}}} 
\left/
\frac{\dot M \sqrt{D}}{10^{-4} \, M_\odot \, \text{
    yr}^{-1}} \right)^{2/3}
\right. 
\end{equation}

with the terminal velocity $\varv_\infty$, the mass-loss rate $\dot{M}$, and the stellar radius $R_\star$ (defined at continuum $\tau_\text{Ross}=20$). We allow for wind inhomogeneities and use the density contrast  $D$ as the 
factor by which the density in the clumps is enhanced compared  to a homogeneous wind of the same $\dot{M}$. We account for wind clumping in the approximation of optically thin structures \citep{Hillier1991,Hamann1998}.
Models with a smooth wind result in a  too strong C\,{\scshape iv} resonance doublet relative to the optical emission lines. A value of $D=4$, as found for massive WN-type stars, improves the fit quality, but higher values of $D$ cannot be excluded.

Our stellar atmosphere models include atomic models for H, He, C, N, O, and Si. Iron-group line blanketing is treated by means of the superlevel approach, i.e. the thousands of energy
levels of the iron group elements (Sc-Ni) are combined to energy bands. Between the energy bands the line transitions for the NLTE calculations are treated as superlines with a finite width and cross section $\sigma(\nu)$, comprising the data for millions of spectral lines \citep{Graefener02}.

\begin{figure}
    \centering
    \includegraphics[width=\columnwidth]{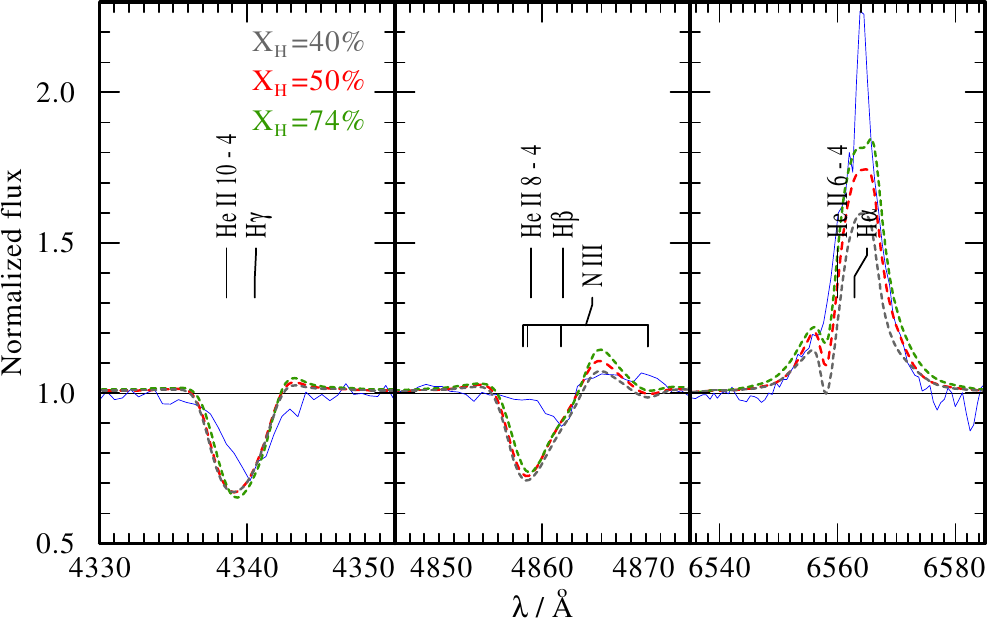}
    \caption{Central star of Hen\,2-108: Details of the optical spectrum, observation (blue solid line) and
    models with different hydrogen abundances (dashed lines, see key)}
    \label{fig:cs-hydrogen}
\end{figure}

While the fit quality to the metal lines is in general acceptable or good, it is worse for the helium lines in the parameter range under consideration. Hence, the helium and hydrogen abundance is only roughly constrained. Moreover, the fit to the Balmer lines is hampered by the blending with stellar He\,{\scshape ii} lines of the Pickering series and contamination with nebular emission, which might not be completely removed. 
The best fit  is for $X_\text{H} =50\%$, but solar hydrogen abundance gives also an acceptable fit (see Fig.~\ref{fig:cs-hydrogen}). 

For silicon and oxygen we obtain a sufficient fit quality with solar abundances, while carbon and nitrogen seem to be slightly enhanced. 
The carbon and nitrogen abundances are inferred from the many C\,{\scshape iii}, C\,{\scshape iv}, N\,{\scshape iii}, and (very weak) N\,{\scshape iv} lines. From the various silicon lines the Si\,{\scshape iv}~$\lambda\lambda$6668-6702 multiplet is quite sensitive to the silicon abundance. For the determination of the oxygen abundance we use the only line in the optical range, O\,{\scshape iii}~$\lambda$5592, which is also quite sensitive.
We notice the pronounced N\,{\sc V} resonance line in the UV that can only be reproduced by our models when super-ionziation by X-ray emission is taken into account. For this purpose, an optically thin hot gas component of $T=2\,$MK is assumed to be distributed within
the stellar wind. We only account for its free-free emission
(thermal bremsstrahlung), as the filling factor is arbitrarily chosen \citep[cf.][for the formalism]{baum1992}.

Finally, we fit the spectral energy distribution (SED) to the observed \textit{IUE} spectra and broadband photometry (see Fig.~\ref{fig:CSsed}) by using the geometric distance from \citet{Bailer-Jones2021} and applying the reddening law of \citet{Cardelli89} with an $R_V=2.0$, which gives a slightly better fit than the standard value of 3.1. We infer an $E_{B-V}=0.37\,$mag and hence determine a stellar luminosity
of $\log (L_\star/L_\odot)=3.62$. From $L_\star$ we also obtain a stellar radius of $R_\star=1.25\,R_\odot$ and a mass-loss rate of  $\log(\dot{M}/(M_\odot\,\text{a}^{-1}))=-7.2$.
Because of the low value C/O=0.22 (Section \ref{sec:phys}), and the CS with H abundance of 0.50$^{+0.24}_{-0.10}$ by mass (Table \ref{tab:csparam}), we suggest the CS of Hen~2-108 is not a result of a very Late Thermal Pulse (VLTP) or a LTP, since the significant H abundance and low C/O ratio in the nebula of ”born-again” stars is not predicted by models \citep{Hajduketal20}.

In the literature, many works have used the weak emission lines stars [WELS] as a placeholder for unknown types of objects. \citet{Tylenda93}, for example, has classified Hen\,2-108 as [WELS] but pointed out that the spectrum was of low quality. Works such as \citet{Weidmann11} concluded that the presence of the carbon emission lines was an indication of a [WC] type Wolf-Rayet, but classified the CS of Hen~2-108 as a [WELS]. \citet{Parthasarathy98} classified the CS of Hen~2-108 as [WC]-PG1159 type, that is, [WELS] in a transition phase between the post-AGB and pre-white dwarf phases. 
We note that in our spectrum, all stellar absorption features in the optical spectrum appear to be wind lines and specifically H$\beta$ has a P Cygni line profile. Therefore and in analogy to the the classification scheme by \citet{crowther2011} for massive stars, one can assign the spectral class [Of/WN] to the CS of Hen\,2-108 with respect to the N\,{\scshape iii}~$\lambda\lambda$4634–41 emission lines and H$\beta$. In contrast to the examples shown in their paper, we do not detect N\,{\scshape v} in the optical range.
The detailed subtype of Hen\,2-108 is [Of/WN8] when applying the classification scheme established by \cite{smith1968} for massive WN stars. For this subtype, the N\,{\scshape iii} lines are much stronger than N\,{\scshape iv} lines and comparable to the He\,{\scshape ii}~$\lambda$4686 line. 
We use brackets to distinguish this object from the massive Of/WN stars.

\section{Photoionization Model} 
\label{sec:structuremodels}

We used the one-dimensional photoionization code \textsc{Cloudy} (version 17.02; \citealt{Cloudy2017_Reference}) and the Python library \textsc{PyCloudy} \citep[version 0.9.6;][]{Morisset_2013} to construct a pseudo-3D photoionization model for Hen~2-108. Our main goal with the model is to produce a better understanding of the results presented in the previous sections and of the characteristics of Hen~2-108, in particular, its matter distribution and ionization structure and ionizing source.

\textsc{Cloudy} nebular simulations provide integrated line fluxes, which we can use to compare with the observations in search for a Hen~2-108 photoionization model. Using \textsc{PyCloudy} and assuming a matter distribution for the nebula, we are able to use \textsc{Cloudy} models to simulate not only integrated fluxes, but also line maps and spatial profiles to compare to our spatially resolved observations. For this model, we will not attempt to reproduce the most central zone of the nebula, as the emission is strongly contaminated by the stellar emission, as seen in previous sections.

The model we construct is based on our observations, the analysis we present in the previous sections, and information available in the literature. Table~\ref{tab:modelA_comp} lists the observed integrated line fluxes and infrared photometric measurements used to constrain the model. Fluxes of some lines are dominated by the central (stellar or nebular) emission and are not used to constrain the model.

We included Spitzer Space Telescope \citep{spitzerWerner04} mid infrared line fluxes from P\citet{Pottasch11} in our analysis.
\citet{Pottasch11} made corrections for aperture effects and these line ratios are suitable for comparison to our observed and modelled flux ratios for the whole nebula. \citet{Pottasch11} provide the H$\beta$ absolute flux of 1.3$\times$10$^{-11}$~erg~cm$^{-2}$~s$^{-1}$ derived from the 2 and 6~cm radio flux obtained by \citet{Milne_Aller_1975,Milne_Aller_1982}. As they pointed out, this flux is consistent with the H$\beta$ flux of 3.7$\times$10$^{-12}$~erg~cm$^{-2}$~s$^{-1}$ measured by \citet{Acker92}, if the extinction coefficient is $c$(H$\beta$)$=$~0.53. For comparison, we obtained similar values of $c$(H$\beta$)$=$~0.52 and $F$(H$\beta$) = 1.9$\times$10$^{-12}$~erg~cm$^{-2}$~s$^{-1}$ for the integrated spectrum. \citet{Pottasch11} suggest an upper limit for the [\ion{C}{iii}]~$\lambda$1909 ultraviolet line flux obtained from IUE (International Ultraviolet Explorer, \citealt{iueBoggess78}). IRAS (Infrared Astronomical Satellite, \citealt{irasNeugebauer84}) fluxes from \citet{2015A&C....10...99A} were used to constrain the dust content in Hen~2-108. As discussed IRAS field of view seems to include all or most the emission from Hen~2-108, no corrections were used when comparing these fluxes with our modelled fluxes. Also, we have no evidence of significant contamination from field objects within the IRAS field of view.

\begin{table} 
    \centering
    \caption{Hen~2-108 modelled nebular fluxes and comparison to observations.}
    \label{tab:modelA_comp}
    \begin{tabular}{lcccc}
    \hline
    Value  & \multicolumn{2}{c}{$I(\lambda)$} & $\kappa(O)$ \\
    
    & Observations & Model & \\
            
    \hline
    \multicolumn{4}{l}{\textbf{Optical/VLT:}} \\
    \ion{H}{i} 6563~\AA\   	& 295.0	    &	280.4	&	-0.28	\\
    \ion{H}{i} 4340~\AA\   	&  43.01	&	47.15	&	0.35	\\
    
    \ion{He}{i} 4471~\AA\  	    &	5.35	&	3.71	&	-0.90	\\
    \ion{He}{i} 5876~\AA\  	    &  16.46	&  10.56	&	-1.69	\\

    \ion{He}{ii} 4686~\AA\  	&	$<<$1.06	&	0.02	&	--	\\
        
    {[}\ion{N}{ii}] 5755~\AA\ 	&	0.62	&	0.71	&	0.33	\\
    {[}\ion{N}{ii}] 6548~\AA\ 	&  25.96	&  28.50	&	0.36	\\
    {[}\ion{N}{ii}] 6583~\AA\ 	&  79.50	&  84.30	&	0.22	\\

    \ion{N}{ii} 5679~\AA\	    &	$<$0.36	&	0.01	&	--	\\

    {[}\ion{O}{ii}] 7320~\AA\ 	&	1.43	&	4.08	&	2.59	\\
    {[}\ion{O}{ii}] 7330~\AA\ 	&	1.26	&	3.34	&	2.40	\\
    
    {[}\ion{O}{iii}] 4363~\AA\  &   0.20	&	0.61	&	2.75	\\
    {[}\ion{O}{iii}] 4959~\AA\  &  55.90	&  57.96	&	0.14	\\
    {[}\ion{O}{iii}] 5007~\AA\  & 168.0	    & 172.9	    &	0.16	\\
    
    \ion{C}{ii} 4267~\AA\ 	    &	0.58	&	2.23	&	 3.32	\\
    \ion{C}{ii} 7231~\AA\ 	    &	0.46	&	0.15	&	-2.76	\\
    \ion{C}{ii} 7236~\AA\ 	    &	0.74	&	0.03	&	-7.91	\\
    
    {[}\ion{S}{ii}] 6716~\AA\ 	&	1.83	&	1.87	&	0.05	\\
    {[}\ion{S}{ii}] 6731~\AA\ 	&	2.40	&	2.55	&	0.15	\\
    
    
    {[}\ion{Ar}{iii}] 7136~\AA\ &	15.71	&  18.86	&	0.70	\\
    
    {[}\ion{Cl}{iii}] 5518~\AA\ &	0.46	&	0.45	&	-0.05	\\
    {[}\ion{Cl}{iii}] 5538~\AA\ &	0.40	&	0.42	&	0.12	\\
    
    {[}\ion{Fe}{iii}] 4658~\AA\ 	&	0.57	&	0.48	&	-0.42	\\
    {[}\ion{Fe}{iii}] 5270~\AA\ 	&	0.23	&	0.41	&	1.43	\\
	\\
    \multicolumn{4}{l}{\textbf{MIR/Spitzer:}}\\
    {[}\ion{Ne}{ii}]  12.8~$\mu$m 	&	102.0	&	96.35	&	-0.19\\
    {[}\ion{Ne}{iii}] 15.5~$\mu$m 	&	16.5	&	28.96	&	1.51\\
    {[}\ion{S}{iii}]  18.7~$\mu$m 	&	72.2	&	3.84    &	--\\
    {[}\ion{S}{iv}]   10.5~$\mu$m	&	3.6	    &	3.70	&	--\\
    {[}\ion{Ar}{iii}]  9.0~$\mu$m 	&	25.5	&	24.47	&	-0.11\\
    {[}\ion{Fe}{iii}] 22.9~$\mu$m	&	1.0	    &	1.43	&	0.71\\
	\\
    \multicolumn{4}{l}{\textbf{MIR Photometry:}} \\
    5 GHz (mJy)     	&	33.00	&	38.90	&	0.41	\\
    100 $\mu$m (Jy) 	    &	7.45	&	4.71	&	-1.13	\\
    60 $\mu$m (Jy)  	&	16.70	&	13.88	&	-0.46	\\
    24 $\mu$m (Jy)  	&	<9.70	&	4.31	&	--	\\
    12 $\mu$m (Jy)  	&	<0.59	&	0.28	&	--	\\
    8.6 $\mu$m (Jy) 	&	0.18	&	0.11	&	-1.11	\\
	\\
    \multicolumn{4}{l}{\textbf{Absolute Flux:}}	\\
	\\
    log F(H$\beta$) 	&	-10.89	&	-10.85	&	0.53 \\
    (erg~cm$^{-2}$~s$^{-1}$) \\    
    
    \hline     
    \end{tabular}
\end{table}

We use the factor $\kappa(O)$,
\begin{equation}
\kappa(O) = \frac{\log(O_\mathrm{mod}) - \log(O_\mathrm{obs})}{\tau(O)},
\label{eq:QF}
\end{equation}

\noindent defined in \citet{2009A&A...507.1517M} as a measure of the proximity of observed and modelled values in Table~\ref{tab:modelA_comp} and, therefore, the quality of the model. In the expression, $O_\mathrm{mod}$ and $O_\mathrm{obs}$ are the observed and modelled values of the observable. The tolerance factor for this observable, $\tau(O)$, is defined as
\begin{equation}
\tau(O) = log\left(1 + \frac{\Delta I}{I}\right)
\label{eq:tau}
\end{equation}

\noindent for any quantity $I$ with uncertainty $\Delta I$. For the optical line fluxes, we assumed the relative uncertainties $\Delta I / I$ to be
\begin{equation}
\frac{\Delta I}{I} = 
    \begin{cases}
        0.5 & \textrm{if } I \leq 0.1H\beta\\
        0.3 & \textrm{if } 0.1H\beta < I < H\beta\\
        0.2 & \textrm{if } I \geq H\beta
    \end{cases}
\end{equation}

\noindent We added 15 percent to the relative uncertainties of the UV and IR line fluxes. For the absolute H$\beta$ flux, we assume $\Delta I / I$ = 0.15, while for the IR band fluxes, we assume $\Delta I / I$ = 0.50. With such values, we take into account the observational uncertainties and the systematic effects that can cause deviations from the model for an observable. Values of $\kappa$(O) between -1 and 1 indicate a good fit for the correspondent quantity.

To better constrain our model, we also compare the spatial profiles of bright lines extracted from the VIMOS maps to the profiles we obtain from the pseudo-3D models. The comparison is only done by visual inspection, as the goal is to reproduce only the general behaviour, given that the geometry we are assuming for the model is simplified. For the comparison, we use the profile extracted in the N-S direction, as this is more symmetrical. The profiles taken in the E-W direction are similar, although somewhat more extended and less symmetrical. This comparison is especially helpful to constrain the nebular matter distribution, the ionic structure, and the nebular size. 

We considered $T_\textrm{eff}$, $L_\textrm{bol}$, density distribution, elemental abundances of He, C, N, O, Ne, S, Cl, Ar, and Fe, dust-to-gas ratio, and PN distance as free parameters to be determined by the modelling procedure.


\begin{table}
    \centering
    \caption{Hen~2-108 photoionization model parameters.}
    \label{tab:params_modelA}
    
    \resizebox{\columnwidth}{!}{%

    \begin{tabular}{lc|lc}
    \hline
Parameter	&	Value	&	Parameter	&	Value	\\
\hline							
\multicolumn{2}{l}{\textbf{Ionizing Source:}}			&	\multicolumn{2}{l}{\textbf{Dust:}}			\\
$T_\textrm{eff}$ [kK] 	&	40.0 	&	Composition 	&	 Graphite 	\\
log ($L_\textrm{bol}/L_\odot$) 	&	3.18  	&	Size distribution 	&	 {\sc Cloudy} ISM	\\
    	&		&	Dust-to-gas ratio 	&	 6.6$\times$10$^{-3}$ 	\\
\textbf{Distance [kpc]:}	&	 4.0 	&	Dust Mass (M$_\odot$)	&	2.3$\times$10$^{-4}$	\\
    	&		&		&		\\
\multicolumn{2}{l}{\textbf{Gas:}}			&	\multicolumn{2}{l}{\textbf{Gas Abundances:}}			\\
Geometry	&	Spherical	&	 He/H 	&	0.125	\\
 $n_\textrm{H}$ 	&	Radial distribution	&	 C/H  	&	 5.0$\times$10$^{-3}$	\\
	&	 as in Fig.~\ref{fig:modeldensity}	&	 N/H  	&	 4.9$\times$10$^{-5}$	\\

 <Ne.Np>$^{(a)}$ [cm$^{-3}$] & 1350   &  O/H  	&	 4.8$\times$10$^{-4}$     \\

Inner Radius [cm]	&	 10$^{15}$	& S/H  	&	 8.0$\times$10$^{-7}$	 	\\
Outer Radius [cm]	&	 3.03$\times$10$^{17}$	&	 Ar/H 	&	 5.1$\times$10$^{-6}$	 	\\
	&	(Ioniz. Bounded)	&	Ne/H 	&	1.7$\times$10$^{-4}$\\
Gas Mass [M$_\odot$]	&	0.19	&	 Cl/H 	&	1.6$\times$10$^{-7}$ 	\\
	&		&	 Fe/H 	&	1.1$\times$10$^{-6}$	\\
    \hline     

\multicolumn{4}{c}{\textbf{$^{(a)}$ Volume average as defined in {\sc Cloudy} v.17.02 documentation.}}\\

    \end{tabular}%
    }
\end{table}

\begin{figure}
\begin{center}
\includegraphics[width=\columnwidth]{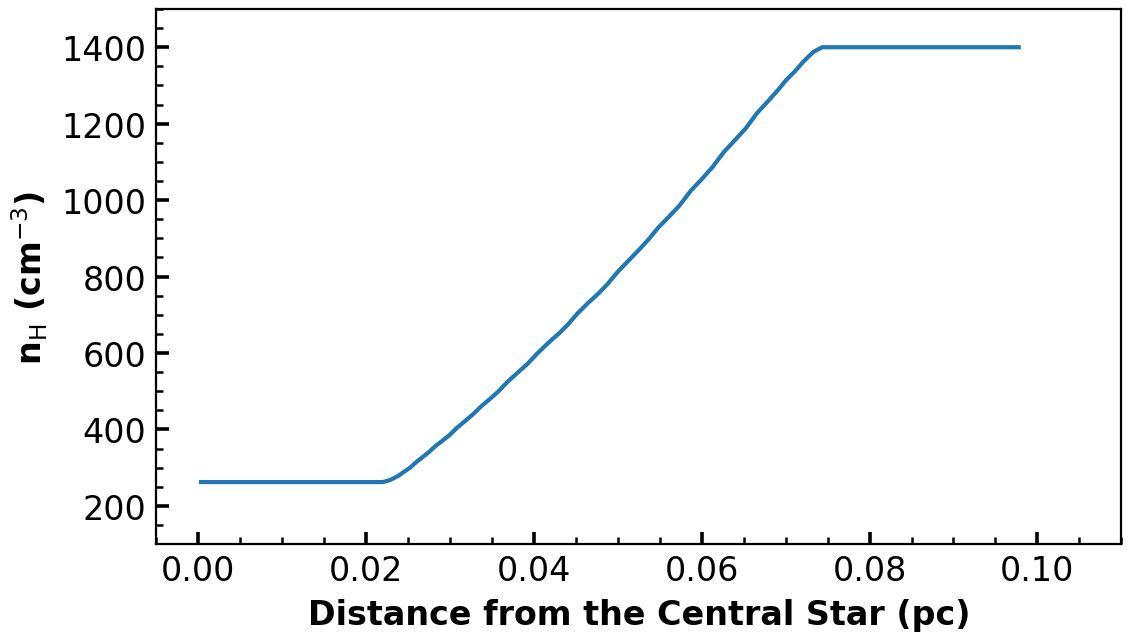} 
\caption{Radial density distribution used in the Hen~2-108 photoionization model.}
\label{fig:modeldensity}
\end{center}
\end{figure}

The parameters of the best photoionization model we found are listed in Table~\ref{tab:params_modelA}. The radial density distribution used is shown in Fig.~\ref{fig:modeldensity}. This model shows a reasonable match to Hen~2-108 integrated and spatially resolved properties. The comparison of the observed integrated fluxes to the corresponding modelled values are given in Table~\ref{tab:modelA_comp}. The model reproduces most of the lines fluxes within the expected errors, in particular the H I, [O III], [N II], [S II], [Ar III] (optical and mid-IR), [Cl III], and [Fe III] (optical and mid-IR) brightest lines.

\begin{figure*}
\begin{center}

\includegraphics[width=5.6cm]{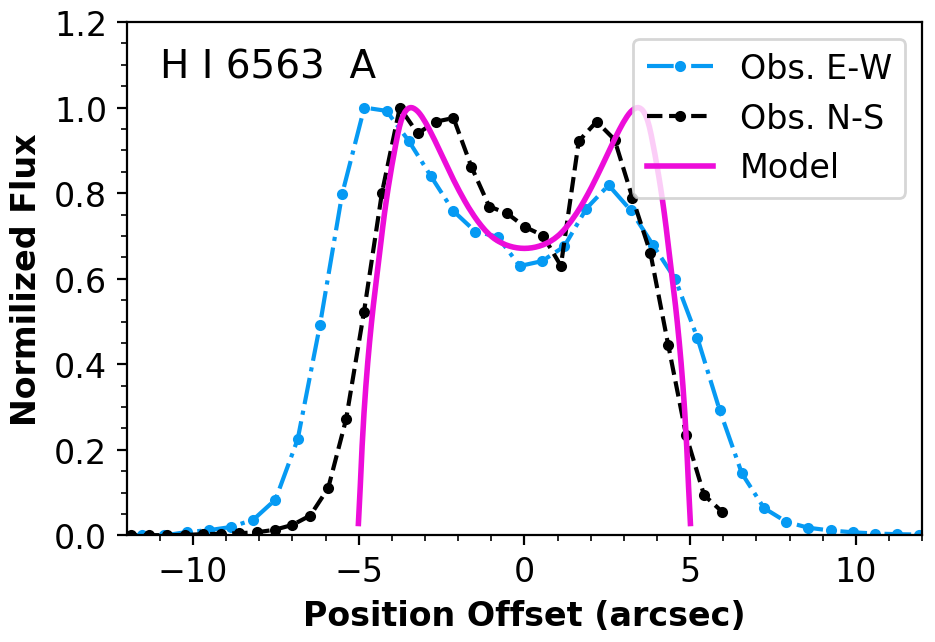} 
\includegraphics[width=5.6cm]{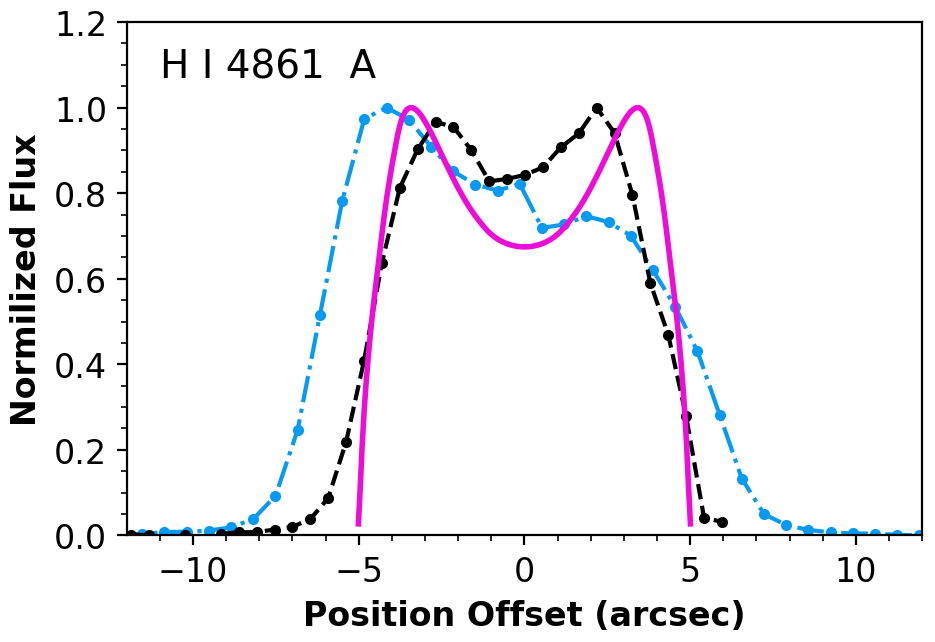} 
\includegraphics[width=5.6cm]{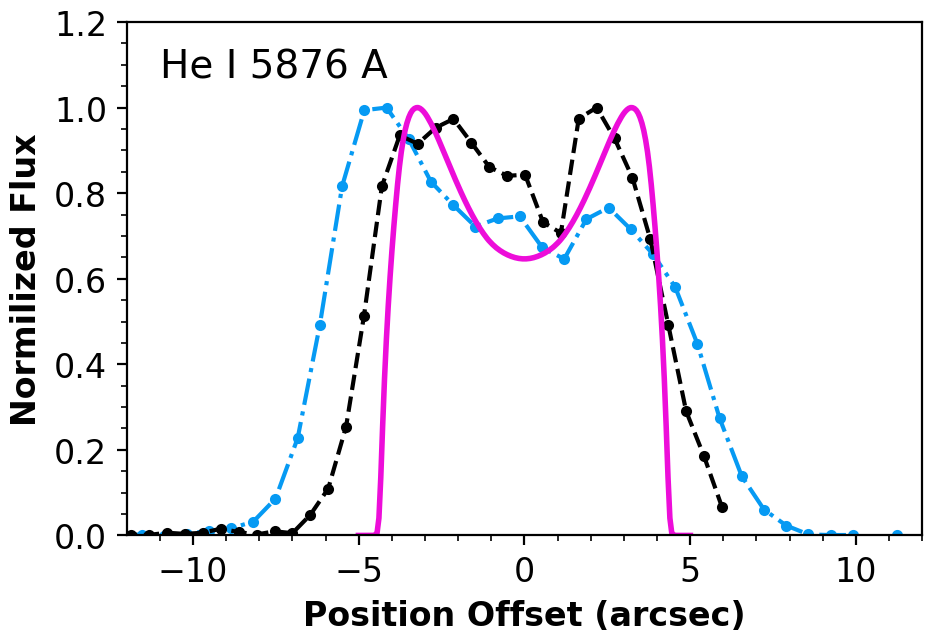} 

\includegraphics[width=5.6cm]{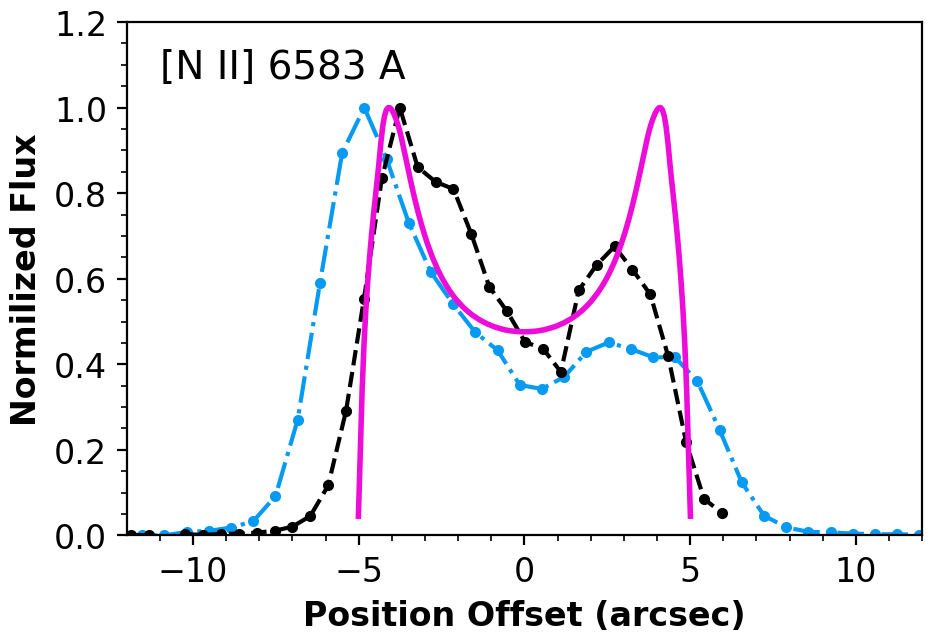}
\includegraphics[width=5.6cm]{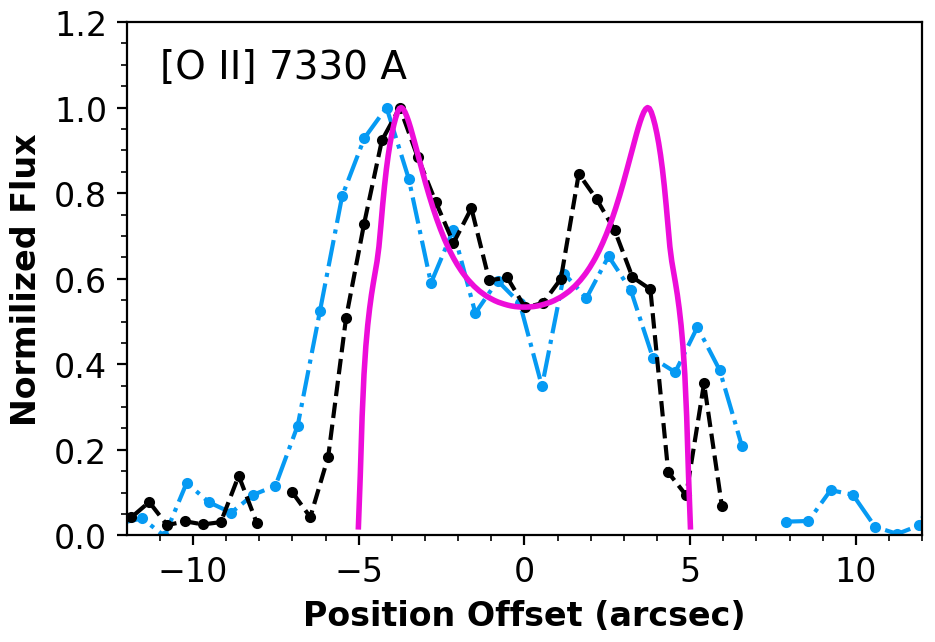}
\includegraphics[width=5.6cm]{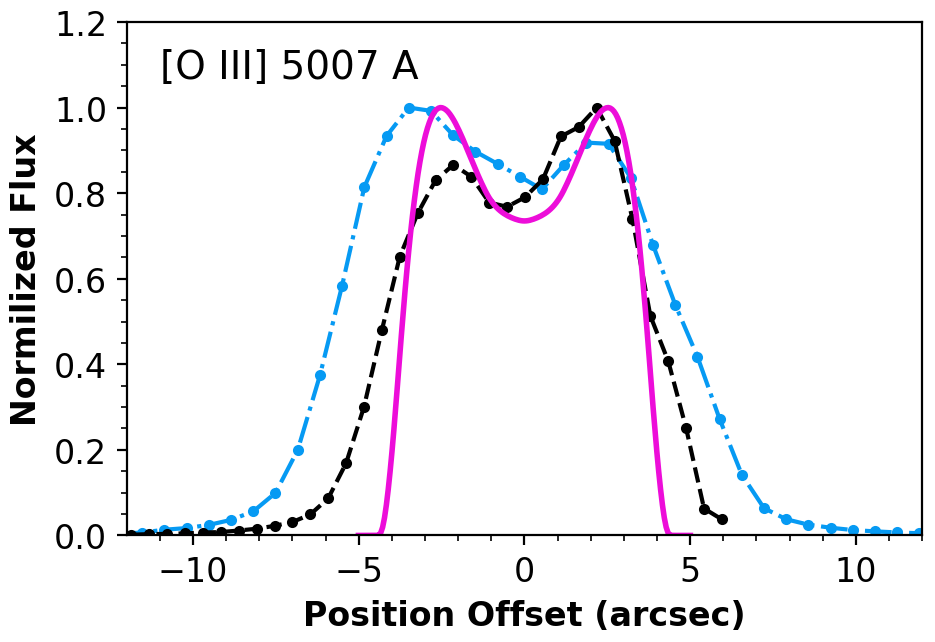}

\includegraphics[width=5.6cm]{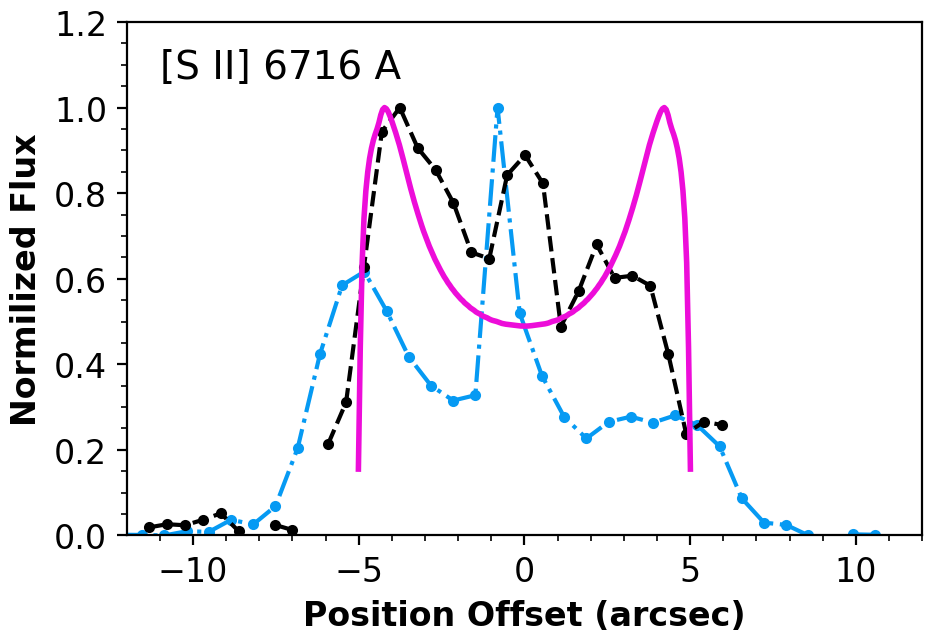}
\includegraphics[width=5.6cm]{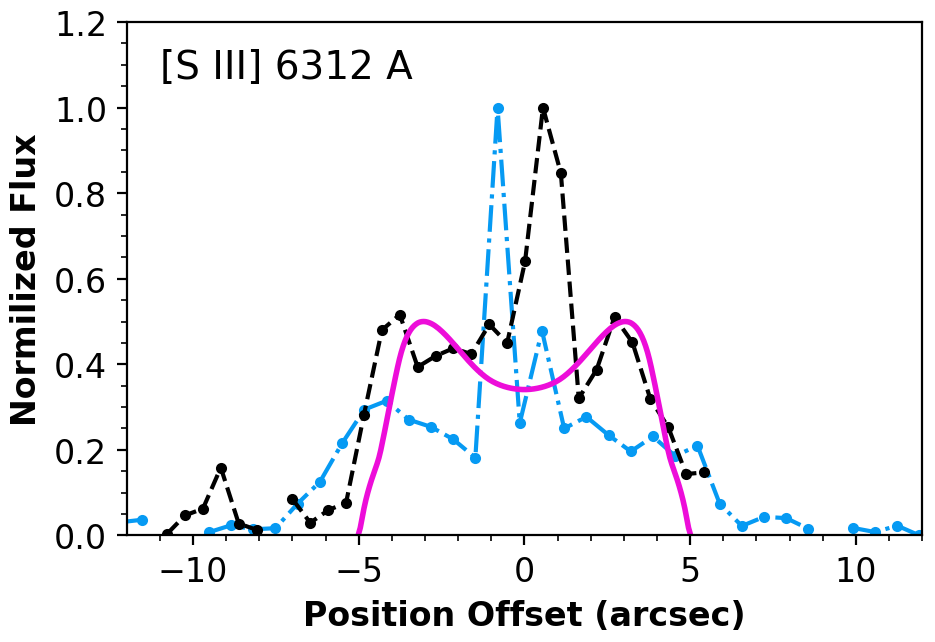}
\includegraphics[width=5.6cm]{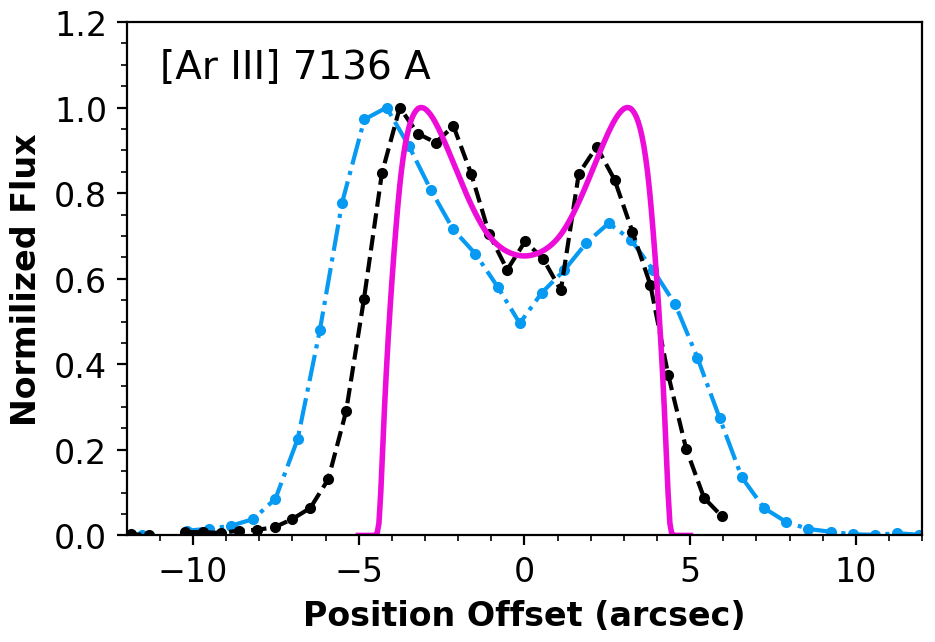}

\caption{Line spatial profiles across the nebula. Black dashed and blue dot-dashed curves are obtained from the emission maps in the N-S and E-W directions (cord passing by the PN centre). The pink solid curve are radial spatial profiles determined from the pseudo-3D model. For the observations, the vertical scale is normalised by the maximum value in each curve. The modelled profiles are scaled to better match the observations (see text). The horizontal scale was aligned according to the centre determined by the \ion{He}{ii} emission peak.}
\label{fig:profiles}
\end{center}
\end{figure*}

Line spatial profiles obtained from the emission maps in the N-S and E-W directions (chord passing by the PN centre) are compared with the radial profiles determined by the {\sc pycloudy} simulation in Fig.~\ref{fig:profiles}. In the figure, the modelled profiles are scaled to better match the normalised observations. In most of the cases, this means normalised to its maximum value as in the observational profiles.

For the simulation, we assumed a spherically symmetrical nebula. This is a simplified geometry, but it provides reasonable matches to the main features for the bright lines radial spatial profiles.

To determine the line flux spatial profiles, we constructed a pseudo-3D photoionization model with spherical symmetry, but with a non-uniform density distribution. We note that Hen~2-108 is not perfectly spherical; it displays a slightly elliptical morphology, as shown in our line maps (see Fig.~\ref{fig:flux108-2}) and inferred by other authors in the literature (e.g. \citealt{Tylenda_etal_2003}). The spherical distribution is a simplification, but including a radial density profile provides a next major step in understanding the PN true 3D structure. Uniform density models could not reproduce important characteristics of the line emission maps of Hen~2-108. For example, in the case of \ion{H}{i}, [\ion{O}{iii}], and [\ion{Ar}{iii}] in, the observed emission in the central regions (Fig.~\ref{fig:flux108-2}) is significantly fainter than it should be in the case of a uniform filled sphere. As Hen~2-108 is a low excitation PN, the central decrease in [\ion{O}{iii}] could not be due to a ionization effect, i.e. the presence of \ion{O}{iv} emission in the central region. In fact, there is no [\ion{O}{iv}]~25.9~$\mu$m emission detected in the Spitzer spectrum of this nebula (see Fig.~\ref{fig:spitzer_spec}). The central decrease is an indication that the density in the central regions of Hen~2-108 must be lower than in the outer zones of the nebula. This could have important consequences for its ionization structure and how well we could use the model to interpret the observed line maps. 

For our models, we explored different radial density distributions that could reproduce a ring structure as seen in the H~I and [O~III] maps. Models with large empty central cavities (i.e., large nebular internal radius, a few pixels wide) do not reproduce very well the mid infrared dust thermal emission. These models lack warm dust emission in the low mid-IR wavelengths. We assume a small nebular internal radius of 10$^{15}$~cm, chosen as to be smaller or around 0.5 pixel in our images, for distances previously estimated for He~2-108 (around 2-5~kpc). The gas density is increased in the radial direction, from a low value in this radius to a large value in the outer radius. The outer radius was limited by the comparison between the modelled and the observed emission line profiles. The best results (match of observation to  were found for the total hydrogen density distribution given in Fig.~\ref{fig:modeldensity}.

Fitting the oxygen and helium lines (fluxes and profiles) simultaneously was challenging. The best global result we got does not produced a very good match between model and observations for all the \ion{He}{i} lines. Typical residuals are of $\sim$35 percent. To improve the \ion{He}{i} line ratios fitting, one possibility would be to increase the helium abundance to a value considerably larger than the value inferred from the empirical analysis. Another possibility would be a matter bounded nebula. As we can see in Fig.~\ref{fig:profiles}, the \ion{He}{i} and \ion{H}{i} lines spatial profiles have similar shapes. For a PN with a low $T_\textrm{eff}$, the He$^+$ region is smaller than the H$^+$ region. They may have similar sizes if the nebula is matter bounded \citep{Osterbrock06}. This was observed, for example, for the PN Tc~1 \citep{Aleman_etal_2019_Tc1}. We also note that the observed regions that emit [\ion{O}{iii}] and [\ion{O}{ii}] in Hen~2-108 have similar outer radius. This could also indicate that the nebula is matter bounded. However, while a matter bounded nebula improve the fitting of the \ion{He}{i} lines to a much better level, the match to observation of other line fluxes and the dust continuum were sacrificed.

A third possible explanation is that the matter distribution is different from what we assume. The emission maps and derived profiles show that the nebula has asymmetries, which are obviously not reproduced by the our spherically symmetric model. The asymmetries are seen in most of the line profiles, but are more evident in the low-ionization emission as [\ion{S}{ii}], [\ion{N}{ii}], and [\ion{O}{ii}] doublets. More detailed 3D modelling is needed to test this hypothesis.

The observed profiles of the \ion{H}{i} and \ion{He}{i} recombination lines show only a small central emission excess, with the exception of \ion{He}{i} $\lambda$4471, which may have a somewhat prominent central peak emission. Other observed profiles with contribution from the central region are \ion{C}{ii} $\lambda$7231, \ion{N}{ii} $\lambda$5679, [\ion{Fe}{iii}] $\lambda\lambda$4658,5270, [\ion{S}{ii}] $\lambda\lambda$6716,6731, [\ion{S}{iii}] $\lambda$6312, and [\ion{Cl}{iii}] $\lambda\lambda$5518,5538. Some of these lines might have contributions from the CS emission, while in other cases the contribution may be from a stellar wind or nebular emission close to the central star. The density diagnostic map in Fig.~\ref{fig:denstemp-He2-108} indicates a high density (up to $\sim$4000~cm$^{-3}$) in the central pixels region (where the excess emission is seen). The [\ion{S}{ii}] lines, which were used to derive the density diagnostic maps, should be nebular. Due to their low critical densities, the [\ion{S}{ii}] lines ratios are not a useful diagnostic for densities higher than 10$^4$~cm$^{-3}$ \citep{Osterbrock06}, but the [\ion{Fe}{iii}] lines indicate the possible presence of a hot dense gas in the core of Hen~2-108 (see Sect.~\ref{sec:phys}). The models we test with such densities, however, did not reproduce the observations. One possible explanation is that the high density region has an open geometry (our models only considered closed geometry) as, for example, a torus, barrel, or a blob.


Distance was assumed as a free parameter determined from the model procedure. The best model distance is 4.0~kpc, which is close to the recent calculations of 3.6$\pm$0.2~kpc by \citet{Bailer-Jones2021} and 4.1~kpc by \citet{2022arXiv220604458A} based on GAIA parallax measurements.





The nebular gas elemental abundances are assumed to be constant across the nebula. The abundances determined from our observations (Table~\ref{tab:diagNP108}) were used as initial guesses. Values were varied to improve the match to observations.

The dust content and composition in Hen~2-108 has not been published in the literature. The C to O abundance ratio in Hen~2-108 is not well constrained as the C abundance is poorly known. Although, Pottasch et al. (2011) reported a possible fullerene line in this nebula, which could indicate the presence of C-rich dust, we have found no evidence of fullerene emission the low-resolution (see Fig.~\ref{fig:spitzer_spec} in Appendix~\ref{ap:spitzer}) or in the high resolution Spitzer spectra of Hen 2-108. Mid infrared bands of polycyclic aromatic hydrocarbons (PAHs) are also not detected in the mid-IR Spitzer spectra of this nebula (Fig.~\ref{fig:spitzer_spec}). However, preliminary models we calculated indicated that the gas should be C-rich and the dust continuum is reasonably fitted with graphite dust. For our models, we assume that the dust grains are composed of graphite with the standard {\sc Cloudy} ISM size distribution \citep[i.e., a power law of slope -3.5, with sizes in the range from 0.005 to 0.25~$\mu$m][]{MRN}. Dust is uniformly mixed with the gas. To determine the best model, we only vary the dust-to-gas ratio, which can be well constrained by the available mid and far infrared photometric measurements. It is not our goal to provide a detailed dust model for Hen~2-108 here, but only to consider the general effect of dust on the radiation processing and gas heating, for which our dust model is adequate.

\begin{figure}
\begin{center}
\includegraphics[width=\columnwidth]{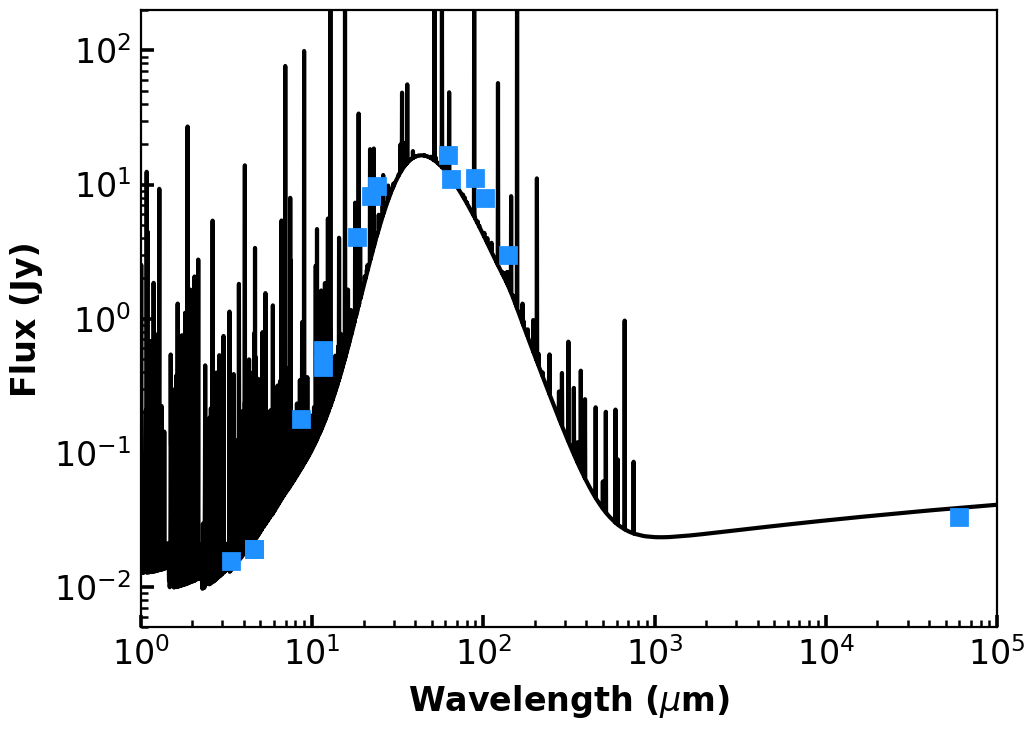} 
\caption{Hen~2-108 simulated spectrum. The blue squares are photometric measurements \citep{2008ApJ...689..194S,Pottasch11,2015A&C....10...99A}.}
\label{fig:model_spectrum}
\end{center}
\end{figure}

The comparison of the infrared continuum simulated with Cloudy and the infrared photometry measurements available in the literature is shown in Fig.~\ref{fig:model_spectrum}. There is a general good match between them. Uncertainties in both model and observations, as well as the contribution of line emission and from field objects may account for the differences. The IR bump shape is reasonably fitted by the model. This indicates that the dust temperature distribution should be close to the real distribution. This is of particular interest as, models with constant density predict that the heated dust close to the CS would produce more near IR emission than we obtained with our model with the tailored density distribution.

The optical line spectrum of Hen~2-108 (Fig.~\ref{fig:integratedspectrum}) shows that this is a low-excitation PN. A few determinations of the CS temperature ($T_\textrm{eff}$) are available in the literature. Values obtained from stellar atmosphere models are in the range of 32 to 39~kK. The Zanstra temperatures of \ion{He}{i} is comparable (26~kK). The \ion{He}{ii} Zanstra temperature, however, shows a much higher value (50~kK), but as we discussed in Section~\ref{sec:CSspec}, the \ion{He}{ii} emission is dominated by the central star emission. We thus discard the temperature derived from this line and explored models with $T_\textrm{eff}$ in the range of 25 to 45~kK. We considered models with a large range of luminosities, from 100 to 10\,000 $L_\odot$, which was progressively constrained in the modelling process.

We assume that the CS emits as a blackbody, which is parametrized by the effective temperature ($T_\textrm{eff}$) and its bolometric luminosity ($L_\star$). This approximation is sufficient to our purposes and final uncertainties expected. Using atmosphere models, for example, shows differences in the line fluxes and derived PN parameters of the order of 5 to 10 percent \citep{Bohigas_2008,Walsh18}. The final model have a central star with $T_\textrm{eff} = 40 kK$ and a luminosity of 1500~$L_\odot$.

The ratio \ion{He}{ii}~$\lambda$4686/H$\beta$ is strongly dependent on $T_\textrm{eff}$. Photoionization models show that the ratio \ion{He}{ii}~$\lambda$4686/H$\beta$ should be more than one order of magnitude smaller than the Hen~2-108 observed value if the nebula is excited by a star with $T_\mathrm{eff} \sim 40$~kK (see Appendix~\ref{ap:He2}). This corroborates the conclusion in Section~\ref{sec:CSspec} that the \ion{He}{ii}~$\lambda$4686 emission is produced by the central star. Our final model predicts a flux of only 0.02 percent of H$\beta$.

As in the case of the \ion{He}{ii} emission, other lines produced dominantly or strongly contaminated by the central source can only be used as an upper limits and were not very useful to constrain the model. Naturally, the model do not reproduce those values as they are not of nebular origin. Examples are the $\lambda$4630 complex, where lines are also complicated by being a strong blend of weak lines, as well as \ion{N}{ii}~$\lambda$5679 and \ion{C}{iii}~$\lambda$5696, which are additionally weak and very uncertain. 
 
The [\ion{S}{iii}] line has also important emission in the central region of the nebula, but it is likely of nebular origin. The central star spectrum extracted from the observations does not show this line and the synthesised spectrum does not predict this line to be significant. Its flux is very uncertain because of noise and contamination by the central star emission, so we did not use this line to constrain the model. Our model predicts that the [\ion{S}{iii}] optical and infrared emission comes from the main nebula as seen in Fig.~\ref{fig:profiles}. The spatial profile agrees very well with the observation for the main nebula (considering the uncertainty in this low-intensity emission). The central [\ion{S}{iii}] emission not explained by the photoionization or the star atmosphere models is evidence that there should be other source of emission within that small region close to the CS.

As we previously mentioned, our model is not expect to reproduce the emission from this region due to the matter distribution limitations. Line and dust continuum can be affected by this extra source. The photoionization model shows that the lines of [S~IV] will be produced dominantly in the central region, while the dust thermal continuum emission in the 12 and 25 $\mu$m bands may have a significant contribution from the central nebular region, which explain why the modelled fluxes exhibit differences from the observations.

The temperature of the gas obtained by the model is in reasonable agreement with the observations given the uncertainties involved. For the N diagnostic ratio, the model and observations agree within 6\%. The O diagnostic ratio however is discrepant with the difference being larger than the uncertainty. The discrepancy can not be explained by a recombination line contribution to the {[}\ion{O}{iii}] 4363~\AA\ or {[}\ion{N}{ii}] 5754~\AA\ lines as the model shows that these are under 2\%. The discrepancy in the O diagnostic may be due to the {[}\ion{O}{iii}] 4363~\AA\ line being very uncertain and a better signal to noise spectrum would be required to better constrain this. The discrepancy may also be due to the simplified structure adopted in the model. 

It is important to note that the high C/H abundance adopted was required to balance the gas heating and keep the gas temperature in a level that explain the emission line fluxes and the dust thermal emission. The abundance is, however, not realistic and should not be taken as an abundance estimate for this element. We explored different density structures, ionising spectra, dust quantities and elemental abundances, among other possibilities to find a solution to this issue. However we were unable to find a combination of parameters that reproduced all the constraints to a good level and also gave a more reasonable C/H abundance. Basically, we need an extra cooling agent to balance the heating from the significant dust quantity in this nebula, which is determined by its infrared dust emission. We opt to increase the C abundance to act as an extra cooling agent, as the C/H is not well constrained by the current available data. Another option would be significantly lowering the dust abundance, which would help the situation by reducing the nebular gas heating. However, this would also significantly reduce the dust infrared thermal emission to a level much lower than the observations shows. In both cases, there would be no change to our main results about the 3D structure of the nebula and the general characteristics of the central star, which are our main goals with this simplified model. Carbon should have an ADF factor of the same order than oxygen. However, as the model does not reproduce well the recombination line fluxes, we cannot estimate the real C/H abundance.

\begin{figure}
    \centering
    \includegraphics[width=\columnwidth]{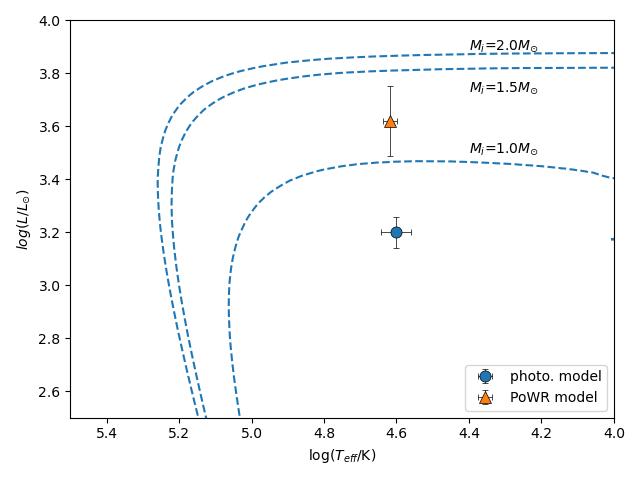}
    \caption{Evolutionary tracks of different progenitor mass for Hen\,2-108. The models for progenitor masses of 1.0, 1.5 and 2.0~M$_{\odot}$ are shown (dashed lines).  } 
    \label{fig:evolution-tracks}
\end{figure}

Finally in Fig. \ref{fig:evolution-tracks} we compare the photoionization model results and the PoWR model results derived in section \ref{sec:CSspec} to theoretical models for the evolution of post-asymptotic giant branch stars and central stars of planetary nebulae with metalicity Z$=0.01$Z$_{\odot}$ from \citet{Bertolami2016}. From this comparison we see that the progenitor mass of the nebula CS is in likely lower than 1.5$M_{\odot}$. The luminosity difference between the PoWR result and that of the photoionization model is probably due to the simplified density structure adopted in the latter and a more detailed 3D model is needed to constrain this further.

\section{Conclusions}
\label{sec:conclusions}

A detailed study of the PN Hen~2-108 was carried out which it allowed us to better understand the characteristics of the nebula and its CS, as well as its evolutionary stage. The data allowed us to analyse the integrated spectrum of the nebula, as well as the spatially resolved emission maps. The integrated spectrum allowed us to obtain measurements from a larger number of emission lines observed compared to previous studies. The emission maps give better details of the general ionization structure of the nebula.

Emission lines of the usual elements were detected, such as H, He, O, N, S, but also lines of C, Cl, Fe and Ar. Among the lines in common with other works, we found a good agreement of the intensities. Abundances were obtained for He, N, O, Ar, S and Cl, which, in general, were consistent with previous results considering the uncertainties, except for N and O, which were higher. For O and N we were able to do the calculation and obtained an ADF(O/H) = $9 \pm 3$ and ADF(N/H) = $22 \pm 15$ respectively. Considerable ADF values such as these have been associated with binary systems in the centre of PNe \citep{Corradi15,Wesson18}. 

The newly developed code {\sc satellite} was also used to analyse the Hen 2-108 VIMOS data set. The {\sc satellite} procedures allowed us to investigate and quantify spatial variations of diagnostics and abundances in 1D and 2D analysis. The angular analysis module revealed a valley in c(\hb) for PAs of the pseudo-slits between 25 and 75~degrees, contrary to $T_e$ and $N_e$ that display no variation. Similarly, the chemical abundances were also found to be unchanged with PAs. On the other hand, the radial analysis module shown that c(\hb) increases as a function of the radius from the CS while $T_e$ and $N_e$ are nearly constant with an augmentation in $T_e$ for radius$<$1.2~arcsec.

Our data also allowed us to obtain the spectrum of the CS, efficiently discounting the nebular emission. With our medium resolution spectrum, we found that all stellar absorption features in the optical spectrum appear to be wind lines and specifically $H\beta$ and He\,{\scshape i}~$\lambda$5876 lines have a P Cygni line profile. The CS spectrum also shows a widening of the H$\alpha$ line, which is characteristic of stars with strong winds resulting from advanced stages of mass loss. The spectrum was then modelled with the Potsdam Wolf-Rayet (PoWR) code for expanding atmosphere to infer the stellar parameters. The detailed model fit allowed us to classify the CS as a star of type [Of/WN8], making it the newest member of a rare group of stars for which only 3 Of-WR(H), 8 [WN] and 11 [WR] are currently confirmed  as listed in \citet{WR-CScat20}. The CS with H abundance of $0.50^{+24}_{-10}$ by mass, together with the low value C/O=0.22 obtained from the nebular abundance analysis suggests that it is not a result of a very Late Thermal Pulse (VLTP) or a LTP, since the significant H abundance and low C/O ratio in the nebula of ”born-again” stars is not predicted by models \citep{Hajduketal20}. Although no specific evolutionary pathway has yet been found for this type of star, it could be the result of AGB final thermal pulse (AFTP) \citep{Werneretal09} or  binary-induced mass exchange or a WD merger as mentioned by \citep{Frew14} for the similar [WN] type. The ADF results we obtain for Hen~2-108, and its connection to binarity would favor the latter scenario, although further studies are needed to confirm this.

We have built a pseudo-3D photoionization model that provided us with a better understanding of the object's matter distribution, ionization structure and the central source. The model was able to reproduce most of the observational constrains available to a reasonable level. The ionizing source is a low effective temperature CS (40~kK) with a luminosity of 1500~L$_\odot$. We note that the temperature of the photoionization model's ionizing source agrees with the WN star spectrum that best reproduces the Hen~2-108 CS spectrum. The photoionization model puts the nebula around 4~kpc from us, in agreement with GAIA parallax data. The model uses spherical symmetry, but the radial matter distribution showed in Fig.~13 was necessary to better fit the observations. The main nebula has a low-density ($n_\textrm{H} = $260~cm$^{-3}$) cavity, with an outer shell with a density of $n_\textrm{H} = $1400~cm$^{-3}$. The total gas mass of the main nebula is 0.19~M$_\odot$ and its size is 0.1~pc. The dust-to-gas mass ratio is 6.6$\times$10$^{-3}$, resulting in a total dust mass of 2.3$\times$10$^{-4}$~M$_\odot$. Asymmetries in the observed profiles with respect to the models indicate that the nebula is not perfectly spherical. Our analysis and modelling efforts also indicates that the emission excess of some lines seen in the central pixels should be produced in an open geometry shell. 
The main limitation of the model is the necessity of a very high C/H abundance to better reproduce the temperature of the gas. The fact that the empirically determined ionic ratios of C and O are of the same order indicates that these two elements should have a similar ADF. However, as the model does not reproduce well the recombination line fluxes, we cannot estimate the real C/H abundance.

Comparing the central source temperature and luminosity obtained from the PoWR and photoionization models to evolutionary tracks of post-asymptotic giant branch stars and central stars of planetary nebulae from \citet{B16}, we can infer that the progenitor star has a mass no larger than 1.5$M_{\odot}$. This result is consistent with the low C/O abundance ratio we find from the observations, which is typical of low mass stars. More detailed high resolution, high signal to noise spectra and three dimensional photoionization models are also required to better constrain the stellar parameters.

\section*{Acknowledgements}
Based on observations collected at the European Southern Observatory under ESO programme 079.D-0117(A). This study was financed in part by the Coordena\c{c}\~{a}o de Aperfei\c{c}oamento de Pessoal de N\'{i}vel Superior - Brasil (CAPES) - Finance Code 001 (B.L.M.M. and I.A.). This work has made use of the computing facilities available at the Laboratory of Computational Astrophysics of the Universidade Federal de Itajub\'{a} (LAC-UNIFEI). The LAC-UNIFEI is maintained with grants from CAPES, CNPq and FAPEMIG. SA acknowledges support under the grant 5077 financed by IAASARS/NOA. This research has made use of the VizieR catalogue access tool, CDS, Strasbourg, France (DOI: 10.26093/cds/vizier). The original description of the VizieR service was published in A\&AS 143, 23. This research has made use of NASA’s Astrophysics Data System.
This publication has benefited from a discussion by a meeting sponsored by the International Space Science Institute (ISSI) at Bern, Switzerland.

\section*{Data Availability}


The VIMOS Hen~2-108 observations are publicly available in the ESO Archive (\url{http://archive.eso.org/eso/eso_archive_main.html}). The photoionization model was generated with publicly available codes and all the parameters necessary to run them are provided. The data sets generated during and/or analysed during the current study are available from the corresponding author on reasonable request.



\bibliographystyle{mnras}
\bibliography{refs} 




\appendix

\section{Spitzer spectra of Hen 2-108} \label{ap:spitzer}

\begin{figure}
\begin{center}
\includegraphics[width=\columnwidth,trim={0.4cm 0.7cm 0 0.0cm},clip]{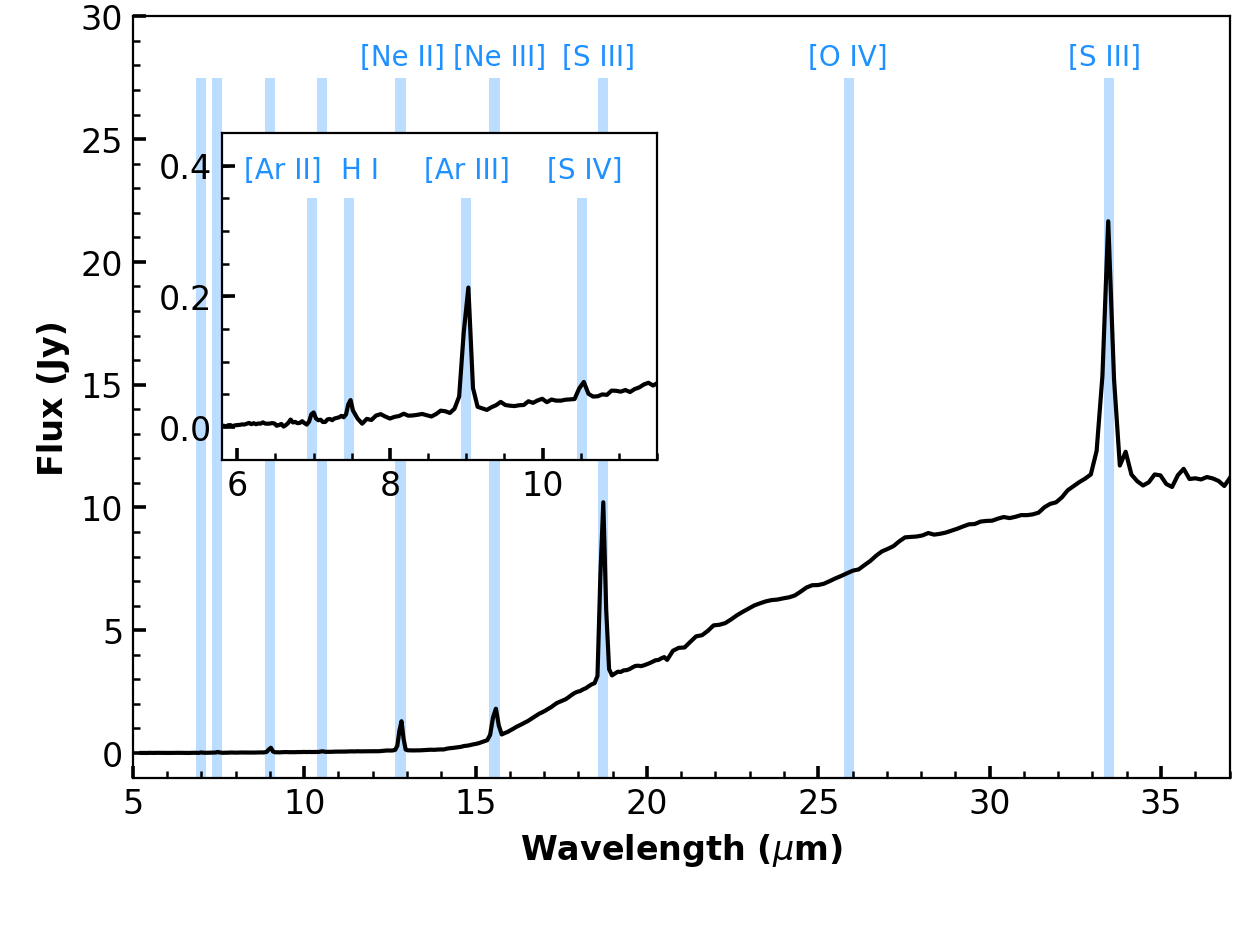}
\caption{Hen~2-108 mid-infrared spectrum obtained from the Cornell Atlas of Spitzer/IRS Sources \citep[CASSIS;][]{cassisLR}. The insert shows a zoom of the low-wavelength spectral region; units are the same as in the main plot.}
\label{fig:spitzer_spec}
\end{center}
\end{figure}

Figure~\ref{fig:spitzer_spec} shows the low-resolution Spitzer Space Telescope mid-infrared spectrum of Hen 2-108. The spectrum was obtained from the Cornell Atlas of Spitzer/IRS Sources \citep[CASSIS;][]{cassisLR}.

\section{He{\sc ii} line dependence with CS temperature} \label{ap:He2}

\begin{figure}
\begin{center}
\includegraphics[width=\columnwidth]{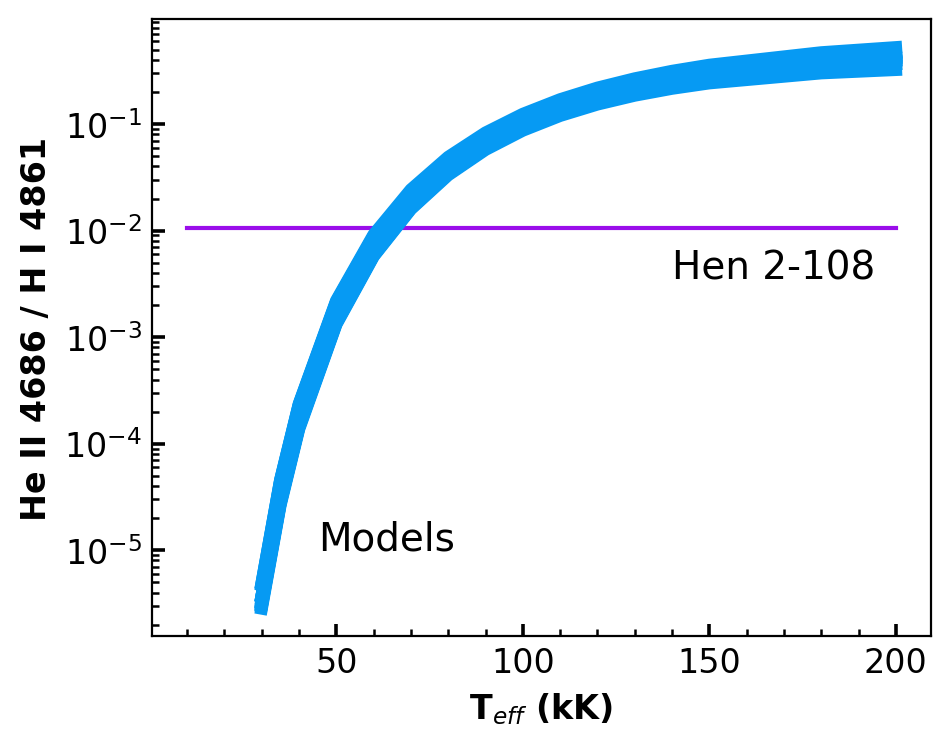}
\caption{The strong dependence of the line ratio \ion{He}{ii} $\lambda$4686 to H$\beta$ with the central star effective temperature. The purple line shows the ratio observed for Hen~2-108. The blue curve represent the modelled values; the curve thickness indicate the variation caused by changing gas density, luminosity and abundances in ranges typical for PNe \citep{Kwok00}.}
\label{fig:he2_line}
\end{center}
\end{figure}

Fig.~\ref{fig:he2_line} shows the \ion{He}{ii}/H$\beta$ intensity ratio strong dependence with $T_\star$
The ratios were calculated for a grid of {\sc Cloudy} models covering a wide range of stellar and nebular parameters, which represents the range observed for PNe. The thickness of the blue curve is due to the superposition of curves for different parameters. The \ion{He}{ii}~$\lambda$4686/H$\beta$ ratio observed for Hen~2-108 ($T_\mathrm{eff} \sim 40$~kK ) should be more than one order of magnitude smaller to be of nebular origin.

To produce Fig.~\ref{fig:he2_line}, we use a grid of {\sc Cloudy} models for $T_\textrm{eff} =$~30-200~kK, $L_\textrm{bol} =$~10$^2$-10$^4$ L$_\odot$, and $n_\textrm{H} =$~10$^2$-10$^5$~cm$^{-3}$. We include models with solar abundances and also models with abundances of the more abundant species varied by a factor of 5 around solar. The models assume a MNR dust distribution, with a dust-to-gas ratio between 0.1 and 10 of that of the ISM (3.3$\times$10$^{-3}$). We only show models with graphite dust, but tests showed that the values in the plots are similar for silicate dust as well. Spherical symmetry is assumed and all the models are ionization bounded. All the models run are plotted in the figure as curves that varies only $T_\mathrm{eff}$, while keeping the other parameters constant.

\bsp	
\label{lastpage}
\end{document}